 \documentclass[aps,pre,twocolumn,groupedaddress, amsmath, amssymb, showpacs]{revtex4-1}

\usepackage{graphicx}
\usepackage{color}

\newcommand{\ave}[1]{\langle #1 \rangle}

\newcommand{\dint}[3]{\int_{#2}^{#3} \! d #1 \,}

\renewcommand{\vec}[1]{\boldsymbol{#1}}

\newcommand{\del}{\vec{\nabla}}

\newcommand{\overbar}[1]{\mkern 1.5mu\overline{\mkern-1.5mu#1\mkern-1.5mu}\mkern 1.5mu}

\newcommand{\ocp}{\mathrm{OCP}}

\newcommand{\lk}{{(l, k)}}
\newcommand{\ordone}[1]{\left[ #1 \right]_1}
\newcommand{\ordtwo}[1]{\left[ #1 \right]_2}
\newcommand{\HD}{$\mathrm H^+ \mathrm D^+$}
\newcommand{\HHe}{$\mathrm H^+ \mathrm{He}^{2+}$}
\newcommand{\HC}{$\mathrm H^+ \mathrm C^{6+}$}
\newcommand{\gmbar}{\overbar\Gamma}

\begin{document}


\title{Effective Potential Theory for Diffusion in Binary Ionic Mixtures}


\author{Nathaniel R. Shaffer$^1$, Scott D. Baalrud$^1$, J\'er\^ome Daligault$^2$}

\affiliation{$^1$Department of Physics and Astronomy, University of Iowa, Iowa City, IA 52242}
\affiliation{$^2$Theoretical Division, Los Alamos National Laboratory, Los Alamos, NM 87545}


\date{\today}

\begin{abstract}

Self-diffusion and interdiffusion coefficients of binary ionic mixtures are evaluated using the Effective Potential Theory (EPT), and the predictions are compared with the results of molecular dynamics simulations.
We find that EPT agrees with molecular dynamics from weak coupling well into the strong coupling regime, which is a similar range of coupling strengths as previously observed in comparisons with the one-component plasma.
Within this range, typical relative errors of approximately 20\% and worst-case relative errors of approximately 40\% are observed.
We also examine the Darken model, which approximates the interdiffusion coefficients based on the self-diffusion coefficients.

\end{abstract}




\maketitle


\section{Introduction}
\label{sec:intro}

Strongly coupled plasmas arise in several frontier topics of physics research. 
In the laboratory, they occur in the imploding fuel capsules of inertial confinement fusion experiments\cite{HuPRL2010}, the highly charged dust of complex plasmas\cite{IkeziPF1986}, and both neutral\cite{KillianPRL1999, StricklerPRX2016} and non-neutral\cite{BollingerPRL1984} ultracold plasmas. 
In nature, strongly coupled plasmas occur in the interior of gas giant planets and white dwarfs, and in the crusts of neutron stars\cite{PaquetteAPJS1986b, BeznogovPRL2013}.
When the plasma consists of more than one ion species, the transport of mass via diffusion becomes a central practical concern.
It is important to optimizing the D-T reaction in inertial fusion efforts, where the different diffusion rates of each isotope and contamination from the plastic shell make optimal mixing a challenge.
In white dwarfs, gravity-driven diffusion influences the chemical composition of the star with substantial ramifications for the evolution of the star and its lifetime estimates, as well as its optical properties\cite{PaquetteAPJS1986b, BeznogovPRL2013}.

The state of the art in quantifying diffusion in strongly coupled plasmas is to conduct molecular dynamics (MD) simulations of systems at thermal equilibrium and extract diffusion coefficients from velocity correlations\cite{HansenPA1985,DaligaultPRL2012, HaxhimaliPRE2014}. 
In order to tabulate the diffusion coefficients of strongly coupled mixtures for use in hydrodynamic models, one must perform MD simulations not only at several coupling strengths, but for each composition of interest. 
Even with modern computing power, the computational cost of such an undertaking becomes impractical if the system spans a range of conditions. 
This underscores the need for a flexible, efficient theory for strongly coupled mixtures. 

Traditional plasma transport theories based on a binary collision picture, e.g., Landau-Spitzer, typically treat only weakly coupled plasmas.
Improvements can be made if one models screening by treating particles as interacting through a Debye-H\"uckel potential\cite{LiboffPF1959}, and extensions to moderate coupling have been proposed via modified screening lengths\cite{PaquetteAPJS1986a}.
Such models, however, show serious inaccuracies at strong coupling\cite{BaalrudPRL2013,DaligaultPRL2016}. 

Effective Potential Theory (EPT) is a recently proposed method for extending Boltzmann-based plasma kinetic theory into the strong coupling regime\cite{BaalrudPRL2013, DaligaultPRL2016, BaalrudPRE2015}.
It relaxes the binary collision assumption by treating particle interactions via the potential of mean force\cite{BaalrudPRL2013}. 
It also treats the excluded volume (or Coulomb hole) in repulsive interactions using a modified version of Enskog's kinetic equation for hard spheres\cite{BaalrudPRE2015}.
In a one-component plasma (OCP), EPT has been shown to accurately predict the self-diffusion coefficient up to coupling strengths of 30\cite{BaalrudPRL2013}, where strong, liquid-like correlations are known to occur\cite{DonkoPRL2002}.
It was recently demonstrated to achieve comparable accuracy for warm dense matter when used in conjunction with the average-atom two-component plasma model\cite{DaligaultPRL2016}.
Beznogov and Yakovlev were the first to use EPT to assess interdiffusion coefficient in binary ionic mixtures (BIMs)\cite{BeznogovPRE2014}. 
The present work builds on theirs by comparing the EPT predictions for both the self-diffusion and interdiffusion coefficients against MD results over several orders of magnitude in coupling strength for a variety of binary ionic mixtures.
It also extends the modified Enskog correction factor to mixtures.

The binary ionic mixture (BIM) is a model plasma consisting of two species of classical positive ions at a temperature $T$, each with a charge $Z_ie$, mass $m_i$, and number density $n_i$, that interact through the Coulomb potential, $v_{ij}(r)=Z_iZ_je^2/r$\cite{HansenPRL1976}.
The electrons are treated as an inert, uniform, neutralizing background.
We estimate the Coulomb coupling strength of the BIM using the parameter
\begin{equation}
  \label{eq:gamma}
  \gmbar \equiv \ave{Z^{\frac{5}{3}}}\ave{Z}^{\frac{1}{3}}\Gamma_0~,
\end{equation}
where
\begin{equation}
  \label{eq:gamma0}
  \Gamma_0 = \frac{e^2}{ak_BT}~,
\end{equation}
and $a$ is the mean inter-ionic spacing given by $n=3/(4\pi a^3)$, $n=n_1+n_2$ is the total ion number density, $k_B$ is the Boltzmann constant, and the angle brackets denote number-weighted averages,
\begin{equation}
  \label{eq:avedef}
  \ave A = x_1A_1 + x_2A_2~,
\end{equation}
with $x_i=n_i/n$ being the mole fraction of each species. We will also refer to the mass fraction of each species,  $y_i=\rho_i/\rho=m_ix_i/\ave{m}$.
When enumerating the species in a BIM, we assign the label ``1'' to the lighter species and ``2'' to the heavier species.

For symmetric mixtures, we find that EPT agrees with MD from weak coupling into the strong coupling regime, up to coupling strengths at which liquid-like behaviors are known to onset.
Typical relative errors of about 20\% are observed. 
The span of coupling strengths where EPT succeeds for mixtures is similar to what was observed for the OCP\cite{BaalrudPRL2013,BaalrudPRE2015}.

The remainder of the paper is organized as follows.
In Section~\ref{sec:ept}, we discuss how to compute diffusion coefficients in the EPT model and extend the modified Enskog correction concept to mixtures.
In Section~\ref{sec:results}, we compare EPT predictions for the diffusion coefficients against MD results.
In Section~\ref{sec:approx}, we  assess the accuracy of the Darken model for the interdiffusion coefficient based on self-diffusion coefficients.

\section{Effective Potential Theory}
\label{sec:ept}
%
\begin{figure}[t]
  \centering
  \includegraphics[width=3in]{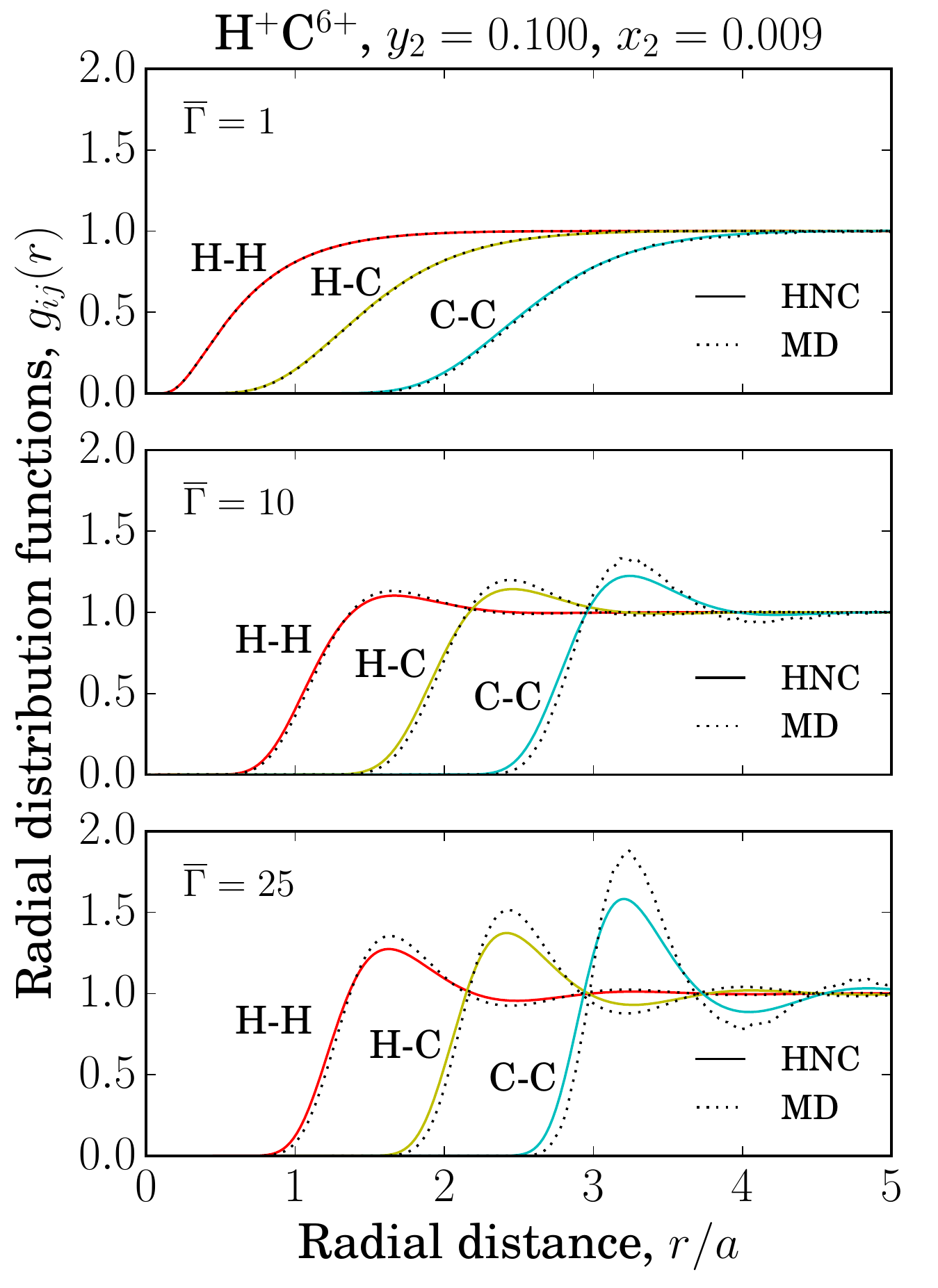}
  \caption{Radial distribution functions in a 10\% carbon (by mass) \HC~BIM\@. Solid lines are hypernetted-chain predictions. Dotted lines are molecular dynamics results.}
 \label{fig:grexample}
\end{figure}
Here, we briefly outline the Effective Potential Theory (EPT) and give detail only when extending on the theory as laid out by refs.~\cite{BaalrudPRL2013,BaalrudPOP2014,BaalrudPRE2015}.
Further details on EPT can be found therein.

In EPT, the transport coefficients are computed using the Chapman-Enskog formulas that result from the solution of the Boltzmann equation with the Enskog correction\cite{ChapmanCowling}.
Instead of using the bare interaction potential to compute the scattering cross-sections for binary collisions, an effective potential is used in order to account for the effect of the surrounding plasma on the mutual interactions between a pair of colliding ions.
In a mixture, the effective potential $\phi_{ij}(r)$ is defined for each pair $(i,j)$ of ionic species.
The effective potential is chosen to be the potential of mean force, which is related to the ion-ion pair distribution function, $g_{ij}(r)$, though\cite{HansenMacDonald}
\begin{equation}
  \label{eq:pmfgr}
  g_{ij}(r) = e^{-\phi_{ij}(r)/k_BT}~.
\end{equation}
The potential of mean force corresponds to the interaction potential between two ions held a distance $r$ apart when the surrounding particles of the plasma are canonically averaged over all configurations.

The effective potential is used to calculate the collision cross-sections for the $\Omega$-integrals that appear in the Chapman-Enskog formulas\cite{ChapmanCowling,FerzigerKaper},
\begin{equation}
  \label{eq:omega}
  \Omega_{ij}^\lk = \sqrt{\frac{k_BT}{2\pi m_{ij}}} \int_0^\infty d\xi~ \xi^{2k+3}e^{-\xi^2}\sigma_{ij}^{(l)}~.
\end{equation}
Above, $m_{ij}$ is the reduced mass, $\xi=|\vec v_i-\vec v_j|/\bar v_{ij}$ is the dimensionless initial relative velocity, $\bar v_{ij}=\sqrt{2k_BT/m_{ij}}$,  and $\sigma^{(l)}$ is the $l$-th binary collision cross-section,
  \begin{equation}
    \label{eq:sigma}
    \sigma^{(l)}_{ij}(\xi) = 2\pi \int_0^{\infty} db~b [ 1 - \cos^l(\pi - 2\Theta_{ij}) ]~,
  \end{equation}
with the scattering angle $\Theta_{ij}$ determined from
\begin{equation}
  \Theta_{ij}(\xi, b) = \int_{r^{\mathrm{min}}_{ij}}^\infty dr~\frac{b}{r^{2}} \left[1 - \frac{b^2}{r^2} - \frac{\phi_{ij}(r)}{k_BT\xi^2} \right]^{-1/2}~, \label{eq:theta}
\end{equation}
where $b$ is the impact parameter and $r_{ij}^{\mathrm{min}}$ is the distance of closest approach.
In practice, it is convenient to also define dimensionless generalized Coulomb logarithms,
\begin{equation}
  \label{eq:xilk}
  \Xi_{ij}^\lk = \chi_{ij} \sqrt{\frac{m_{ij}}{2\pi k_BT}} \left(\frac{2k_BT}{Z_iZ_je^2}\right)^2 \Omega_{ij}^\lk~,
\end{equation}
where $\chi_{ij}$ is a modified Enskog correction obtained from the pair distribution functions, discussed further below.

To evaluate $\Xi_{ij}^\lk$, we require only the ion-ion pair distribution functions, $g_{ij}(r)$, which we obtain by solving the Ornstein-Zernike relation with the hypernetted-chain (HNC) closure,
\begin{align}
  & \hat{h}_{ij}(k) = \hat{c}_{ij}(k) - \sum_s n_s \hat{h}_{is}(k)\hat{c}_{sj}(k)  \label{eq:oz} \\ 
  & g_{ij}(r) = e^{-\frac{v_{ij}(r)}{k_BT} + h_{ij}(r) - c_{ij}(r)} \label{eq:hnc}~,
\end{align}
where $h_{ij}(r)=g_{ij}(r)-1$ and hats denote Fourier transforms\cite{HansenMacDonald}.
Sample solutions for $g_{ij}(r)$ are plotted with MD results in Figure~\ref{fig:grexample}, showing that HNC is accurate up to the development of liquid-like correlations at strong coupling.

\begin{figure}[b]
  \includegraphics[width=3in]{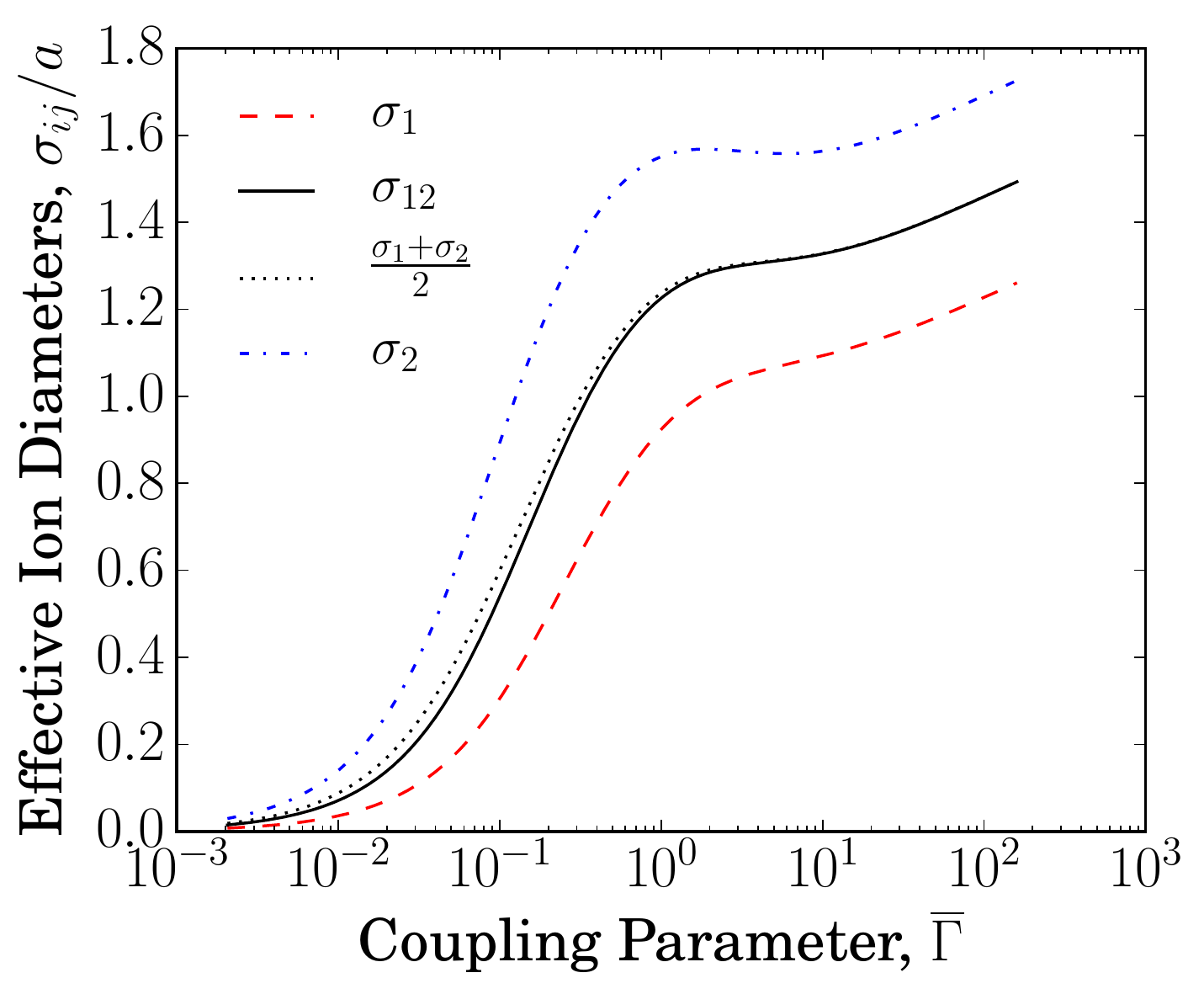}
  \includegraphics[width=3in]{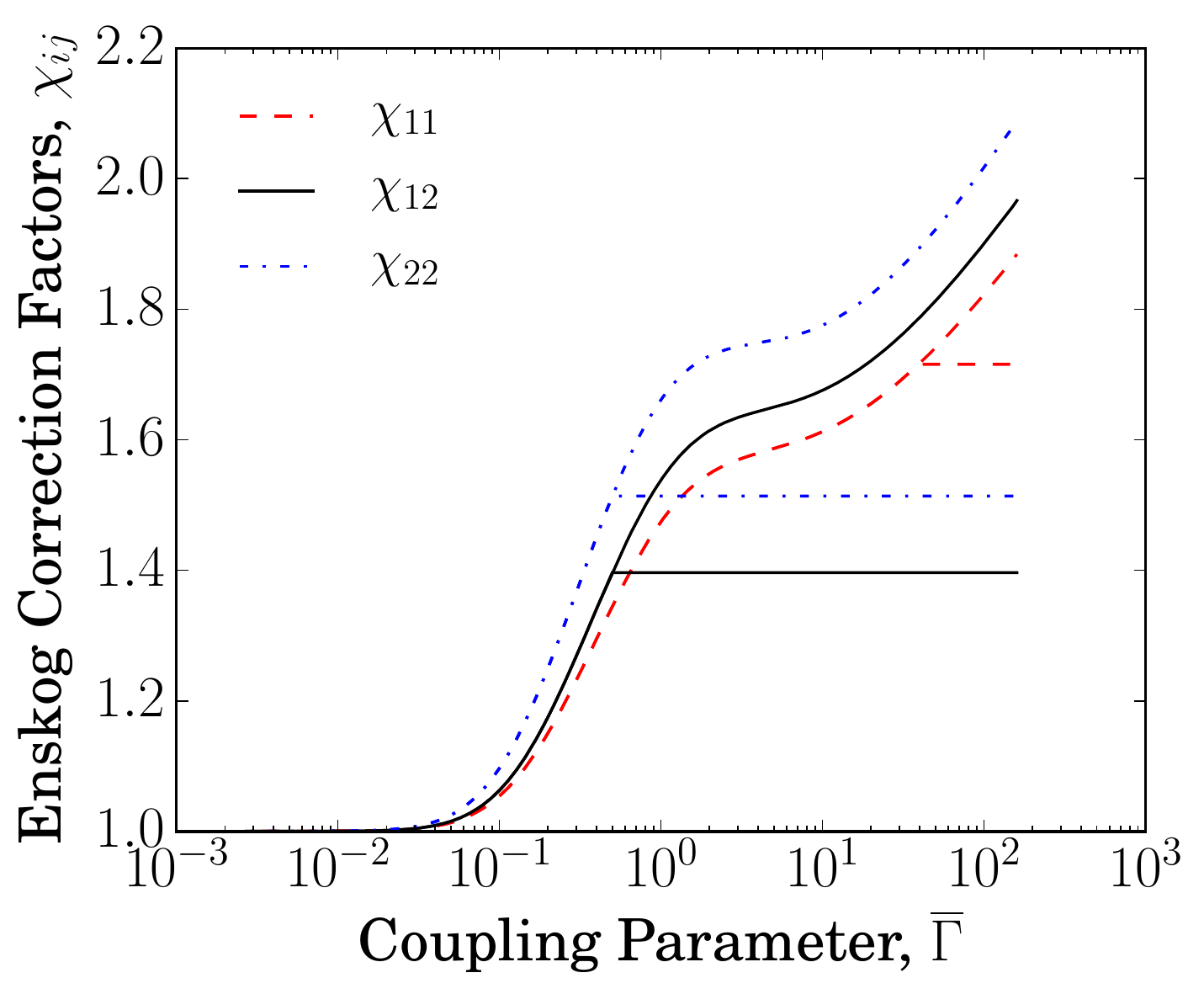}
  \caption{(a) Effective ion diameters and (b) modified Enskog correction factors (with and without virial cutoff) in a 36.8\% helium (by mole) \HHe BIM\@.}
  \label{fig:HHe_y2_700_enskog}
\end{figure}

The prefactors $\chi_{ij}$ in Eq.~\eqref{eq:xilk} are correction factors that arise in Enskog's theory of hard-sphere gases\cite{EnskogKSVAH1922}.
They model the increase in collision frequency that occurs when one accounts for the fact that hard spheres collide when their edges make contact, rather than their centers as in the Boltzmann equation.
Extending the OCP work of Ref.~\cite{BaalrudPRE2015} to BIMs, we define three effective hard-sphere diameters, $\sigma_{ij}$, such that
\begin{equation}
  \label{eq:gijsigij}
  g_{ij}(\sigma_{ij}) = 0.87~.
\end{equation}
In evaluating the modified Enskog correction factors, we assume that $\sigma_{12} = \frac{1}{2}(\sigma_{1} + \sigma_{2})$.
This exact property of hard spheres is only approximate in a BIM, but we find it is quite accurate for $\gmbar > 1$, as shown in Figures~\ref{fig:HHe_y2_700_enskog}a and~\ref{fig:HC_y2_100_enskog}a.

To compute $\chi_{ij}$ from $\sigma_{ij}$ requires three relations.
Two such relations can be obtained by requiring that the partial pressures
\begin{equation}
  \label{eq:partial-p}
  \frac{p_i}{n_ik_BT} = 1 + \sum_j\frac{2\pi}{3}\sigma_{ij}^3n_j\chi_{ij}
\end{equation}
coincide with those of the known virial hard-sphere equation of state up to some order in the density.
For the third relation, we follow Pi\~na, who required that the diffusion force also be consistent with irreversible thermodynamics\cite{PinaJSP1974}.
The resultant Enskog correction factors are:
\begin{align}
  &\chi_{11} = 1 + n_1\frac{B_{111}}{B_{11}} + n_2\frac{B_{112}}{B_{11}} \frac{3\sigma_2}{\sigma_1+2\sigma_2}  ~,\label{eq:chi11} \\
  &\chi_{12} = 1 +  n_1\frac{B_{112}}{B_{12}} \frac{3\sigma_{12}}{\sigma_1+2\sigma_2} + n_2\frac{B_{221}}{B_{12}} \frac{3\sigma_{12}}{\sigma_2+2\sigma_1} ~,\label{eq:chi12} \\
  &\chi_{22} = 1 +  n_2\frac{B_{222}}{B_{22}} + n_1\frac{B_{221}}{B_{22}} \frac{3\sigma_1}{2\sigma_2+\sigma_1} ~,\label{eq:chi22}
\end{align}
where
\begin{align}
  & B_{ij} = \frac{2\pi}{3}\sigma_{ij}^3~,\\
  & B_{iij} = \frac{\pi^2}{108} \sigma_i^3 (\sigma_i^3 + 6\sigma_i^2\sigma_j + 15\sigma_i\sigma_j^2 + 8\sigma_j^3) ~,
\end{align}
are hard-sphere virial coefficients.
In Eqs.~\eqref{eq:chi11}-\eqref{eq:chi22}, we have taken Pi\~na's $\alpha_i$ to be equal to $\sigma_i$, corresponding to the choice that the Enskog correction be evaluated at the point of contact.
Other formulations for $\chi_{ij}(\sigma_{ij})$ have been proposed, some based on virial expansions with other choices of the third closing relation (e.g., Ref.~\footnote{Section 16.6 of Ref.~\cite{ChapmanCowling}}) and others on the Percus-Yevick equation (e.g., Ref.~\cite{KincaidJCP1983}).

In our investigations, it was found that the ion diameters determined by Eq.~\eqref{eq:gijsigij} can become very large in certain strongly coupled mixtures.
This is especially so in mixtures containing a high-$Z$ impurity such as the \HC BIM of Figures~\ref{fig:grexample} and \ref{fig:HC_y2_100_enskog}.
The resultant $\chi_{12}$ and $\chi_{22}$ are large enough to cause EPT to significantly underestimate the values of the impurity self-diffusion coefficient and the interdiffusion coefficient compared to MD results, seen in Figures~\ref{fig:bigfig}c and f.
To improve agreement with MD, we developed the following procedure to cut off the value of $\chi_{ij}$ based on the partial-pressures virial expansion, Eq.~\eqref{eq:partial-p}.

Because Eqs.~\eqref{eq:chi11}-\eqref{eq:chi22} are based on an $\mathcal O(n^2)$ truncation of the hard-sphere equation of state, we require that each term in the truncated partial virial expansions be smaller than the previous,
\begin{equation}
  \label{eq:p-ineq}
  1 > \sum_jn_jB_{ij} > \sum_jn_jB_{ij}(\chi_{ij}-1)\qquad i=1,2~.
\end{equation}
For a given BIM composition, we use Eqs.~\eqref{eq:chi11} and \eqref{eq:chi22} unmodified up to the critical coupling strengths, $\gmbar_{*,i}$, for which Eq.~\eqref{eq:p-ineq} is violated (for $i=1,2$ respectively).
We then cut off the value of each $\chi_{ii}$ according to
\begin{equation}
  \label{eq:chi-cut}
  \chi_{ii} \to \min{\left[\chi_{ii}, \chi_{ii}(\gmbar_{*,i})\right]}~.
\end{equation}
The cutoff value of $\chi_{12}$ is chosen by requiring that both species' partial virial expansions obey Eq.~\eqref{eq:p-ineq}; in other words it cuts off beyond $\gmbar \ge \min\left( \gmbar_{*,1}, \gmbar_{*,2}  \right)$.

Cutoff procedures based on limiting the partial packing fractions, $\eta_i=x_i(\sigma_i/2a)^3$, or total packing fraction, $\eta=\sum_i\eta_i$, were also considered but proved ineffective. The packing fractions were modest ($\eta \lesssim 0.35$) in all cases studied here. Further, in the motivating \HC BIM, $\eta_2\sim 0.05$. This leads us to believe that the apparent failure of the Enskog correction in high-$Z$ impurities is not due to close-packing.

Example plots of $\chi_{ij}$ are shown in Figures~\ref{fig:HHe_y2_700_enskog}b and~ \ref{fig:HC_y2_100_enskog}b.
The uncut Enskog corrections tend to be small (less than 10\%) at weak coupling, but their values grow rapidly as the Coulomb hole widens with increased coupling strength (see, e.g., Figure~\ref{fig:grexample}).
The cutoff activates at weaker coupling for larger ions as intended.
One result of this is that $\chi_{12}$ becomes the smallest of the Enskog corrections at strong coupling.

Below we give the expressions for interdiffusion and self-diffusion coefficients in the so-called first- and second-order Chapman-Enskog approximation.
We recall that the level of approximation refers to the number of terms retained in the Sonine polynomial expansion of certain unknown functions occurring in the Chapman-Enskog solution to the Boltzmann equation.

\subsection{Interdiffusion Coefficient}
\label{sec:interdiff}

\begin{figure}[t]
  \includegraphics[width=3in]{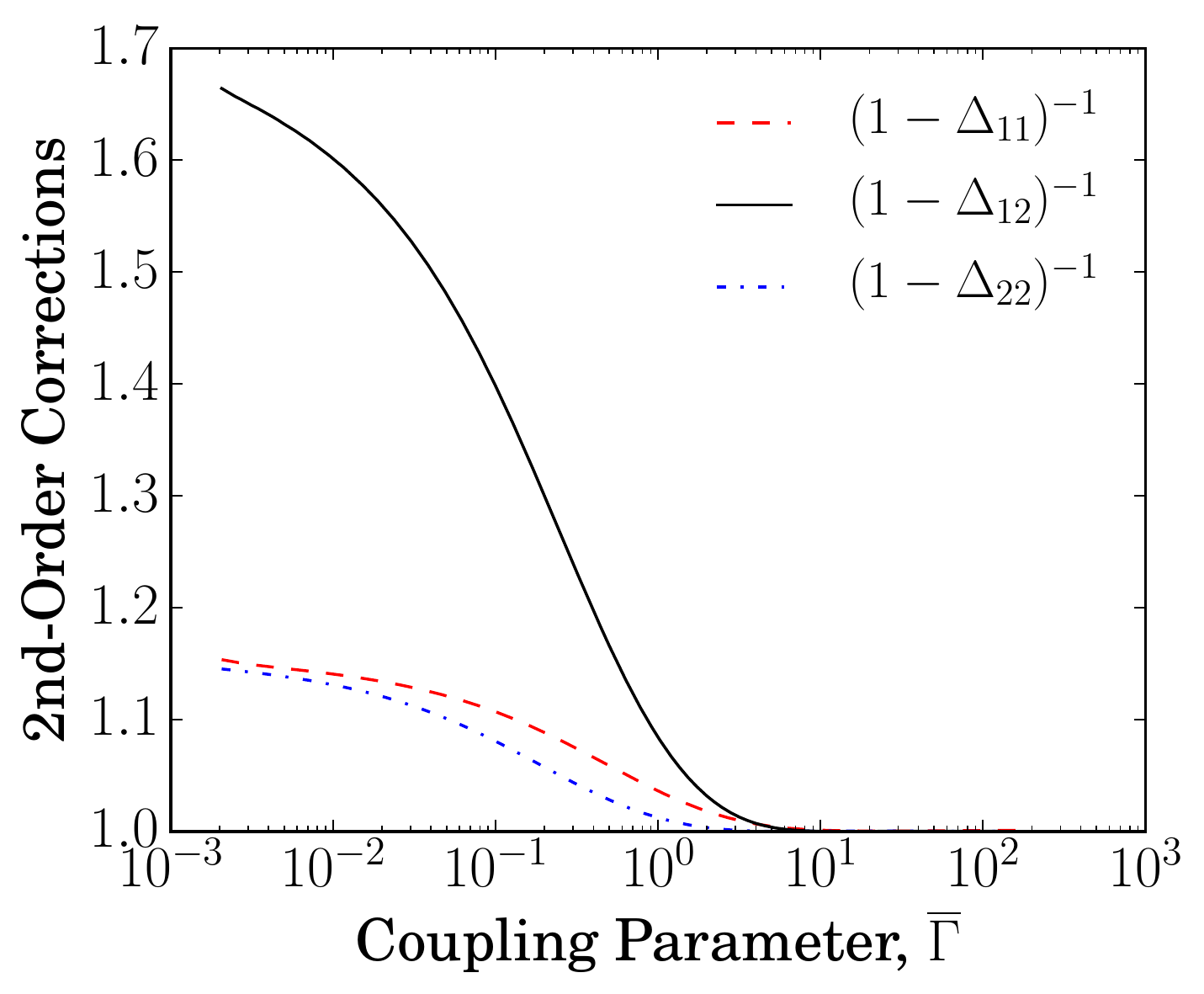}
  \includegraphics[width=3in]{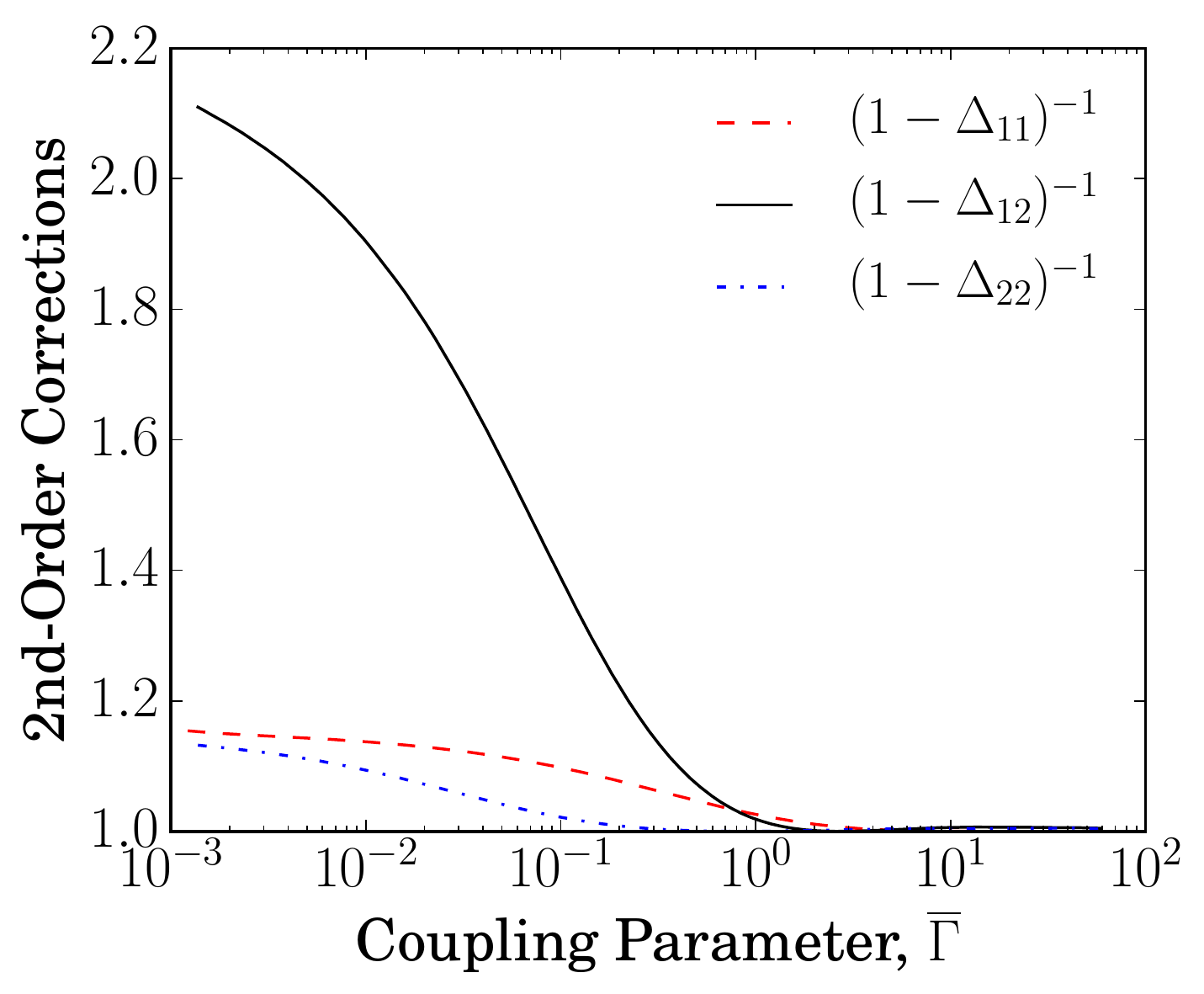}
  \caption{Second-order correction factors for (a) 70\% helium (by mass) \HHe BIM\@ and (b) 10\% carbon (by mass) \HC BIM.}
  \label{fig:ord2}
\end{figure}

In the first Chapman-Enskog approximation, the interdiffusion coefficient is $\ordone{\mathfrak D_{12}}$ is given by\cite{ChapmanCowling}
\begin{equation}
  \ordone{\mathfrak D_{ij}} = \frac{3k_BT}{16nm_{ij}\chi_{ij}}\frac{1}{\Omega_{ij}^{(1,1)}} \label{eq:d12o1}
\end{equation}
The second approximation can be written in the form
\begin{equation}
  \label{eq:d12}
  \ordtwo{\mathfrak D_{ij}} = \frac{\ordone{\mathfrak D_{ij}}}{1-\Delta_{ij}}~,
\end{equation}
where $\Delta_{ij}$ are given in terms of $\Xi_{ij}^\lk$ in Appendix~\ref{sec:deltaij}.
Sample second-order corrections to the interdiffusion coefficient are plotted as solid black lines in Figure~\ref{fig:ord2}a.
They become negligible at strong coupling, where the Sonine polynomial expansion converges rapidly.
%

\subsection{Self-Diffusion Coefficient}
\label{sec:selfdiff}

Coefficients of self-diffusion are defined in terms of single-particle fluctuations at equilibrium, not the collective motion in a fluid.
As such, they do not immediately arise in Chapman-Enskog transport theory, since it gives only the collective transport properties.
In order to derive Chapman-Enskog formulas for the self-diffusion coefficients of a binary mixture, it is necessary to express them in terms of interdiffusion coefficients.
To do so, we will split the second component into two equimolar populations that are physically identical yet distinguishable by some non-mechanical means, e.g., color.
By then writing the interdiffusion coefficients of the ternary mixture in microscopic Green-Kubo form and taking the limit where species 2 and 3 are mechanically identical, we can derive formulas for the self-diffusion coefficients in terms of interdiffusion coefficients.

The three-component Maxwell-Stefan equations,\cite{ChapmanCowling,KrishnaIECR2005}
\begin{equation}
  \label{eq:ms3cp}
  \vec d_a = -\sum_{b=1\ne a}^3 \frac{\vec J_{ab}}{\mathfrak D_{ab}} \quad (a=1,2,3)~,
\end{equation}
can be inverted in terms of new coefficients $L_{ab}(=L_{ba})$,
\begin{equation}
  \label{eq:msinv}
  \vec V_a = -\sum_{b=1}^3 \frac{L_{ab}}{x_ax_b}\vec d_b~,
\end{equation}
where $\vec d_a = x_a\del\mu_a/kT$ is the diffusion driving force from species $a$, $\vec J_{ab}=x_ax_b(\vec V_a - \vec V_b)$ is the particle current, $\vec V_a=N_a^{-1}\sum_{i=1}^{N_a}\vec v_i^{a}$ is the drift velocity of species $a$, and $\vec v_i^a$ is the velocity of the $i$-th particle of species $a$.

The $L_{ab}$ of Eq.~\eqref{eq:msinv} can be written in microscopic Green-Kubo form and expanded\cite{KrishnaIECR2005},
\begin{align}
  L_{ab} &= \frac{1}{3N} \sum_{i=1}^{N_a}\sum_{j=1}^{N_b}\int_0^\infty dt \ave{\vec v_i^a(t) \cdot \vec v_j^b(0)} \\
         &= \left\{
           \begin{array}{ll}
             x_ax_bN C_{ab} & a \ne b \\
             x_a \mathfrak D_a + x_a^2N C_{aa}^\star & a = b
           \end{array}
           \right.~,
\end{align}
where 
\begin{align}
  \mathfrak D_{a} &= \frac{1}{N_a} \sum_{i=1}^{N_a}\frac{1}{3}\int_0^\infty dt \ave{\vec v_i^a(t) \cdot \vec v_i^a(0)}~, \label{eq:selfd} \\
  C^\star_{aa} &= \frac{1}{N_a^2}\sum_{i=1}^{N_a}\sum_{j=1\ne i}^{N_a} \frac{1}{3}\int_0^\infty dt \ave{\vec v_i^a(t) \cdot \vec v_j^a(0)}~, \label{eq:cstar} \\ 
  C_{ab} &= \frac{1}{N_aN_b}\sum_{i=1}^{N_a}\sum_{j=1}^{N_b}\frac{1}{3}\int_0^\infty dt \ave{\vec v_i^a(t) \cdot \vec v_j^b(0)}~, \label{eq:cab}
\end{align}
are respectively the self-diffusion coefficient, the cross-correlation function between particles of the same species, and the cross-correlation function between different-species particles\cite{KrishnaIECR2005}.
Here, the angle brackets denote an average over equilibrium configurations.

The interdiffusion coefficients, $\mathfrak D_{ab}$, can be written in terms of $L_{ab}$ by Eqs.~(4-5, 20-21) of Ref.~\cite{KrishnaIECR2005}.
When species 2 and 3 are mechanically identical and equally abundant, $\mathfrak D_2=\mathfrak D_3$ and $C^\star_{22}=C^\star_{33}=C_{23}$, and the interdiffusion coefficients simplify to
\begin{align}
  & \mathfrak D_{12} = \mathfrak D_{13} = (1-x_1)\mathfrak D_1 + x_1 \mathfrak D_2 + x_1 (1-x_1)\tilde C \\
  & \mathfrak D_{23} = \mathfrak D_2 \frac{(1-x_1)\mathfrak D_1 + x_1 \mathfrak D_2 + x_1(1-x_1)\tilde C}{\mathfrak D_1 + x_1\tilde C}~,
\end{align}
where $\tilde C=N\left(C_{11}^\star - 2C_{12} + C_{22}^\star\right)$.
The cross-correlations in $\mathfrak D_{23}$ can be eliminated in terms of $\mathfrak D_{12}$ and the two self-diffusion coefficients.
Rearranging for $\mathfrak D_2$ then gives
\begin{equation}
  \label{eq:d2}
  \mathfrak D_2 = \frac{\mathfrak D_{12} \mathfrak D_{23}}{(1-x_1)\mathfrak D_{12} + x_1\mathfrak D_{23}}~.
\end{equation}

The Maxwell-Stefan equations, Eq.~\eqref{eq:ms3cp}, can be also obtained from Chapman-Enskog kinetic theory.
One may then write in correspondence with Eq.~\eqref{eq:d2} that the first-order self-diffusion coefficient is
\begin{equation}
  \label{eq:d2ce}
  \ordone{\mathfrak D_2} = \frac{\ordone{\mathfrak D_{12}}\ordone{ \mathfrak D_{22}}}{(1-x_1)\ordone{\mathfrak D_{12}} + x_1\ordone{\mathfrak D_{22}}}~,
\end{equation}
where the fact that $\ordone{\mathfrak D_{23}} = \ordone{\mathfrak D_{22}}$ has been used.
Eq.~\eqref{eq:d2ce} generalizes to both species and to second order as
\begin{equation}
  \label{eq:dice}
  \frac{1}{\ordtwo{\mathfrak D_i}} = \sum_{j=1}^{2}\frac{x_j}{\ordtwo{\mathfrak D_{ij}}}~,
\end{equation}
where $\ordtwo{\mathfrak D_{ij}}$ are the same as in Eq.~\eqref{eq:d12}.
It should be emphasized that $\ordtwo{\mathfrak D_i}\ne\ordtwo{\mathfrak D_{ii}}$, in contrast to the OCP case.
This can be understood by viewing Eq.~\eqref{eq:dice} through the lens of binary collision frequencies, $\nu_{ij} \sim \ordtwo{\mathfrak D_{ij}}^{-1}$.
The timescale for self-diffusion is set by the total collision frequency $\nu_i=\sum_jx_j\nu_{ij}$, as one would expect from an elementary mean-free-path treatment of self-diffusion\cite{Reif}.

\section{Results and Discussion}
\label{sec:results}

%
\begin{figure*}[t]
  \centering
  \includegraphics[width=\textwidth]{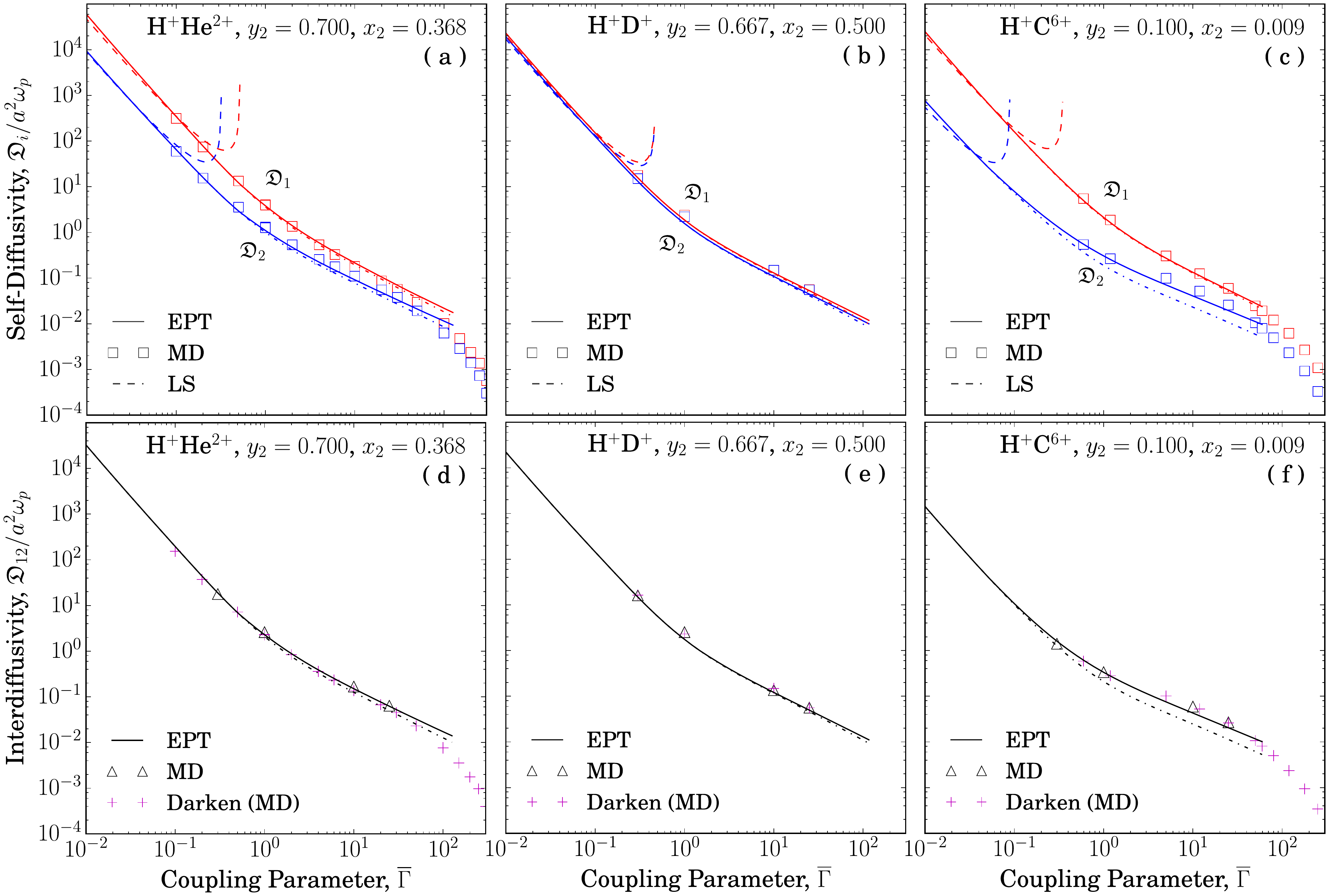}
  \caption{Diffusion coefficients versus BIM coupling parameter, $\gmbar$. (a-c): Self-diffusion coefficients, $\mathfrak D_i$. Solid lines are second-order Effective Potential Theory, shapes are molecular dynamics results, and dotted lines are Landau-Spitzer theory. Dash-dotted lines are EPT without the Enskog correction cutoff. (d-f): Interdiffusion coefficients, $\mathfrak D_{12}$. Solid lines are second-order Effective Potential Theory, shapes are molecular dynamics, and crosses are the Darken approximate using MD self-diffusion coefficients as discussed in Section~\ref{sec:approx}. Dash-dotted lines are EPT without the Enskog correction cutoff. }
  \label{fig:bigfig}
\end{figure*}
We organize our findings into three categories of special interest defined below: ordinary mixtures, equal-$Z$ mixtures, and mixtures containing a high-$Z$ impurity.
In Figure~\ref{fig:bigfig}, we plot the $\gmbar$ dependence of the self- and interdiffusion coefficients of example BIMs that are illustrative of each category.
Each panel of Figure~\ref{fig:bigfig} compares EPT predictions to MD results.

The MD simulations were conducted as described in Refs.\cite{DaligaultPRL2012, JeromeDraft}, and the self- and interdiffusion coefficients were evaluated according to Eq.~\eqref{eq:selfd} and
\begin{align}
  & \mathfrak{D}_{12} = \frac{1}{3Nx_1x_2} \dint{t}{0}{\infty} \ave{\vec{J}_{12}(t) \cdot \vec{J}_{12}(0)}\label{eq:gkinter}~,
\end{align}
respectively.

The EPT self- and interdiffusion coefficients were computed to second order from Eq.~\eqref{eq:dice} and \eqref{eq:d12}, respectively, using the dimensionless form for $\ordone{\mathfrak D_{ij}^{\mathrm{B}}}$ ,
\begin{equation}
  \label{eq:dija2wp}
  \frac{\ordone{\mathfrak D_{ij}^\mathrm{B}}}{a^2\omega_p} =  \frac{\sqrt{\ave{m}/2m_{ij}}}{Z_i^2Z_j^2\ave Z} \frac{\sqrt{\pi/3}}{\Gamma_0^{5/2}}\frac{1}{\Xi^{(1,1)}_{ij}}~,
\end{equation}
where $\omega_p=\sqrt{4\pi n \ave{Z}^2e^2/\ave m}$ is an aggregate plasma frequency.
The EPT diffusion coefficients use the cut off $\chi_{ij}$ described in Section~\ref{sec:ept}, unless explicitly stated otherwise.

In Figures~\ref{fig:bigfig}a-c, we plot the Landau-Spitzer prediction for the self-diffusion coefficients obtained by evaluating Eq.~\eqref{eq:dija2wp} with the substitution
\begin{equation}
  \label{eq:lssub}
  \Xi^{(1,1)}_{ij} \to \log{\Lambda_{ij}} = \log{\frac{1}{\sqrt{3Z_i^2Z_j^2\ave{Z^2}\Gamma_0^3}}}~.
\end{equation}
In Figures~\ref{fig:bigfig}d-f, we include Darken approximation for the interdiffusion coefficient in terms of the self-diffusion coefficients, further discussed in Section~\ref{sec:approx}.

Lastly, we note that both the EPT and MD formulas for the interdiffusion diffusion coefficient exclude the thermodynamic prefactor that arises in the formal theory of diffusion from irreversible thermodynamics\cite{deGrootMazur}.
A complete transport coefficient for use in hydrodynamics applications can be obtained by multiplying Eq.~\eqref{eq:d12} or \eqref{eq:gkinter} by the thermodynamic factor\cite{BoerckerPRA1987},
\begin{equation}
  \mathcal J=\lim_{k \to 0}x_1x_2/S_{xx}(k)~,
\end{equation}
where $S_{xx}(k)$ is the concentration structure factor, which can be calculated directly from the radial distribution functions.
This prefactor enters into the EPT\cite{KincaidPLA1978,LopezdeHaroJCP1983} and MD\cite{HansenPA1985} diffusion coefficients in the same manner, thus its exclusion does not affect the comparison between EPT predictions and MD.

\subsection{Ordinary Mixtures}
\label{sec:ord}
%
\begin{figure}[h]
  \centering
  \includegraphics[width=3in]{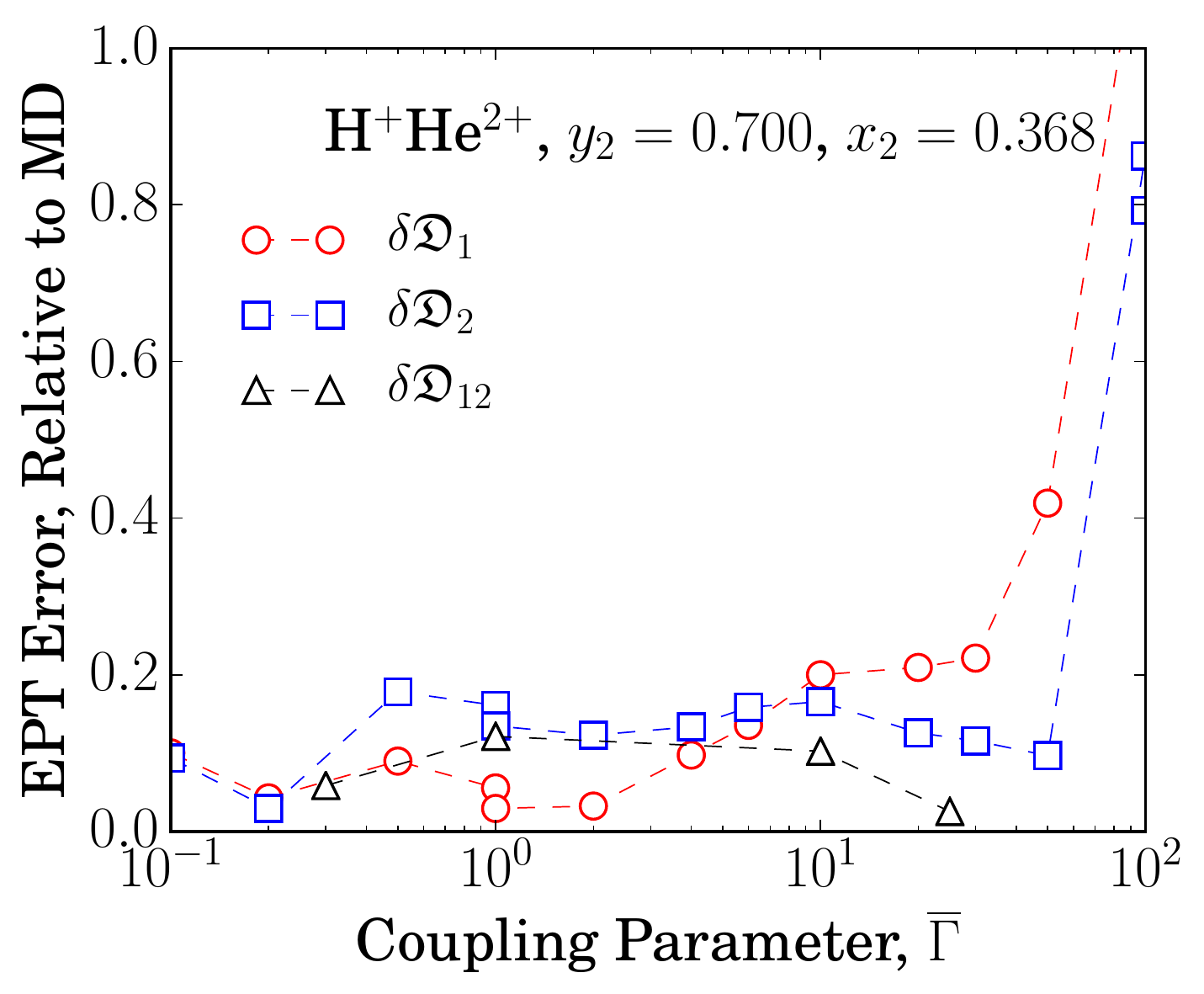}
  \caption{Errors in EPT diffusion coefficients, relative to MD values for a 70\% helium (by mass) \HHe BIM\@. Red circles: hydrogen self-diffusion coefficient. Blue squares: helium self-diffusion coefficient. Black triangles: interdiffusion coefficient. Dashed lines are included to guide the eye.}
  \label{fig:HHe_y2_700_err}
\end{figure}
We refer to an ``ordinary'' mixture as one for which there are no obvious simplifications to be made based on the composition.
That is, there are neither any strong symmetries nor asymmetries that would represent limiting cases of the BIM model.
Diffusion coefficients for an example \HHe BIM are plotted in Figure~\ref{fig:bigfig}a~and~d. 
These exhibit the same qualitative dependence on $\gmbar$ as in the OCP\@. 
In the limit of weak coupling, one can take $\phi_{ij}$ to be a Coulomb potential cut off at the Debye length, in which case the generalized Coulomb logarithms become the traditional Coulomb logarithms\cite{BaalrudPOP2014}, 
\begin{equation}
  \label{eq:coullog}
  \Xi_{ij}^{(1, 1)} =  \log \Lambda_{ij} + \mathcal{O}(1)~,
\end{equation}
and EPT recovers LS theory.

Relative errors between EPT and MD are plotted in Figure~\ref{fig:HHe_y2_700_err}, where the error is 
\begin{equation}
\delta\mathfrak D = |\mathfrak D^{\mathrm{EPT}}-\mathfrak D^{\mathrm{MD}}|/\mathfrak D^{\mathrm{MD}}~.  
\end{equation}
EPT predicts all three diffusion coefficients within $25\%$ of the MD results up to $\gmbar=30$.
Above $\gmbar=30$, the MD diffusion coefficients steeply decrease with increasing $\gmbar$, and the self- and interdiffusion coefficients tend to converge towards a common value.
These trends are associated with the onset of the liquid-like dynamics in the plasma\cite{DonkoPRL2002, DaligaultPRL2012, JeromeDraft}.
In this regime, the binary collision picture on which the EPT relies becomes invalid.

\subsection{Equal-$Z$ Mixtures}
\label{sec:eqz}
%
\begin{figure}
  \centering
  \includegraphics[width=3in]{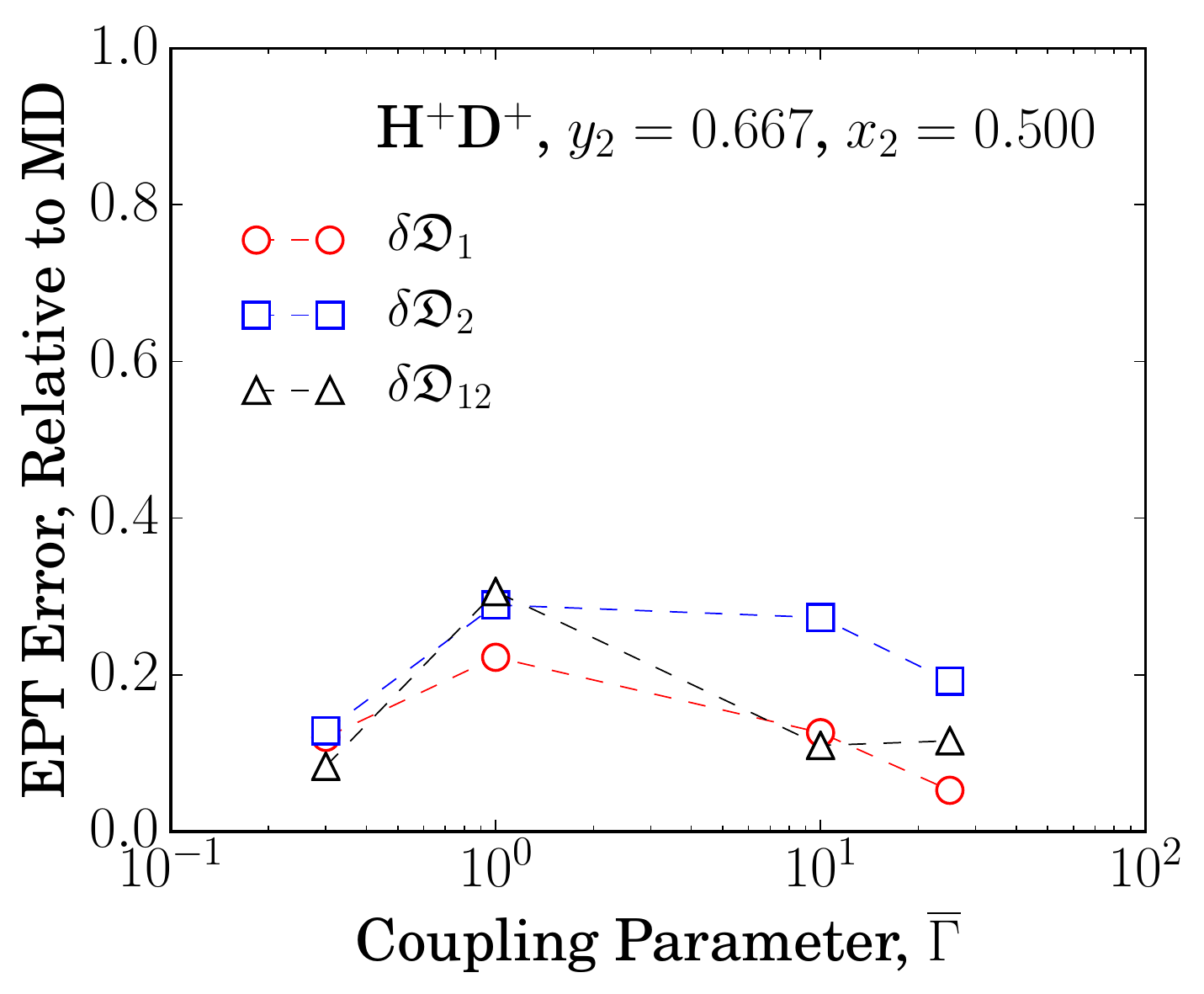}
  \caption{Errors in EPT diffusion coefficients, relative to MD values for a 50-50 (by mole) \HD BIM\@. Red circles: hydrogen self-diffusion coefficient. Blue squares: deuterium self-diffusion coefficient. Black triangles: interdiffusion coefficient. Dashed lines are included to guide the eye.}
  \label{fig:HD_err}
\end{figure}
\begin{figure*}[th]
  \centering
  \includegraphics[width=\textwidth]{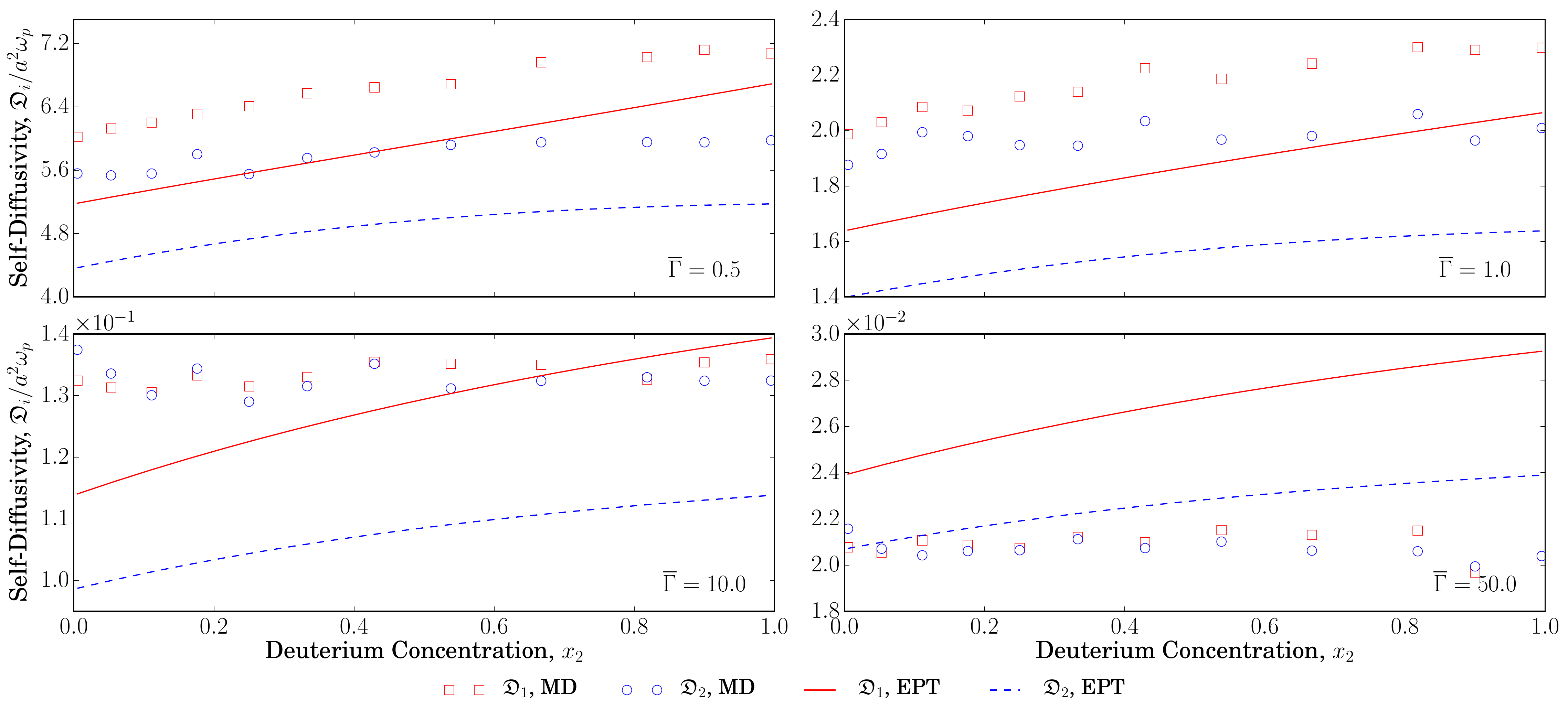}
  \caption{Dependence of the diffusion coefficients on deuterium mole fraction at selected coupling strengths for a \HD BIM\@. Lines are EPT predictions for $\mathfrak D_1$ (solid red) and $\mathfrak D_2$ (dashed blue). Shapes are MD results.}
  \label{fig:HD_vs_x2}
\end{figure*}
In mixtures of equal charge, $Z$, but different masses (e.g., isotopic mixtures), the potentials of mean force are all identical to each other, and they are furthermore the same as those of an OCP with $\Gamma_\ocp = Z^2\Gamma_0$.
It then follows from Eq.~\eqref{eq:xilk} that $\Xi_{ij}^\lk = \Xi_\ocp^\lk$, and the diffusion coefficients are all given according to 
\begin{equation}
  \label{eq:d1ocp}
  \frac{\ordone{\mathfrak{D}_{ij}}}{a^2\omega_p} = \sqrt{\frac{\ave{m}}{2m_{ij}}} \frac{\ordone{\mathfrak{D}_\ocp}}{a^2\omega_p}~,
\end{equation}
where 
\begin{equation}
  \label{eq:docp}
  \frac{\ordone{\mathfrak{D}_\ocp}}{a^2\omega_p} = \frac{\sqrt{\pi/3}}{\Gamma_\ocp^{5/2}}\frac{1}{\Xi^{(1, 1)}_\ocp}
\end{equation}
is the first-order self-diffusion coefficient of an OCP\@.
The second-order corrections depend on the mass ratio in a more complicated way and do not lend themselves to easy comparison with OCP expressions. 
Nevertheless, the ability to use OCP $\phi_{ij}(r)$ data to describe BIM diffusion coefficient for arbitrary masses and concentrations is convenient. 

Relative errors between EPT and MD for a 50-50 \HD mixture are plotted in Figure~\ref{fig:HD_err}.
For the coupling strengths studied, EPT lies within 35\% of the MD values for all three diffusion coefficients.

Due to their simplicity, equal-$Z$ mixtures are useful for isolating and assessing the effect of electronic screening.
To demonstrate this, we consider a simple Debye-H\"uckel potential for the bare ion-ion interaction, 
\begin{equation}
  \label{eq:yukawa}
  v_{ij}(r) = \frac{Z_iZ_je^2}{r}e^{-\kappa_e r}~,
\end{equation}
where $\kappa_e$ sets the strength of electron screening.
Screening weakens ion-ion interactions on average, leading to weaker spatial correlations than in an identical, unscreened plasma. 
At weak coupling, this manifests as an enhancement to the ordinary ion-ion Debye screening. 
At strong coupling, the principal effect is shallower peaks and troughs in $g_{ij}(r)$, resulting in fewer many-body scattering events on average. 
In both cases, one should expect the ions in a screened plasma to be more mobile than those in an unscreened plasma, corresponding to larger values of the diffusion coefficients. 
Both MD and EPT bear out this expectation; a summary is given in Table~\ref{tab:hd_kappa}. 
We note that the effect becomes less pronounced as $\gmbar$ increases. 
From an HNC-EPT perspective, this can be explained by the relatively weak dependence of the $g_{ij}(r)$ peak heights on $\kappa_e$ at strong coupling. 

In Figure~\ref{fig:HD_vs_x2}, we show how the self-diffusion coefficients in a hydrogen-deuterium BIM depend on the relative concentration of the isotopes.
We find that up to $\gmbar \sim 1$, EPT consistently underestimates the MD self-diffusion coefficients, although the trends are captured well.
Beyond $\gmbar\sim 1$, EPT does not capture the convergence of the self-diffusion coefficients to a common value seen in MD; however, Figures~\ref{fig:bigfig}b~and~4e show that the interdiffusion coefficient can still be well-predicted in absolute terms up to $\gmbar=25$.
\begin{table}[h]
  \centering
  \begin{tabular}{c|cc|cc|cc|cc}
    \hline\hline
    $\gmbar$ & \multicolumn{2}{c}{0.5} & \multicolumn{2}{c}{1} & \multicolumn{2}{c}{10} & \multicolumn{2}{c}{50} \\ \hline
    $\kappa_ea$ & 0 & 1 & 0 & 1 & 0 & 1 & 0 & 1 \\ \hline
    $\mathfrak D_1^{\mathrm{EPT}}$ 
             & 6.00 & 8.45 & 1.89 & 2.54 & 0.131 & 0.145 & 0.0274 & 0.0304 \\
    $\mathfrak D_1^{\mathrm{MD}}$ 
             & 6.69 & 8.91 & 2.19 & 2.64 & 0.135 & 0.147 & 0.0215 & 0.0244 \\
    $\mathfrak D_2^{\mathrm{EPT}}$ 
             & 5.00 & 7.05 & 1.58 & 2.13 & 0.109 & 0.122 & 0.0229 & 0.0255 \\
    $\mathfrak D_2^{\mathrm{MD}}$                                   
             & 5.92 & 7.14 & 1.97 & 2.31 & 0.131 & 0.145 & 0.0210 & 0.0237 \\ \hline\hline
  \end{tabular}
  \caption{Self-diffusion coefficients in a 53.8\% deuterium \HD BIM at selected coupling strengths. 
    At each $\gmbar$, unscreened ($\kappa_e=0$) and screened ($\kappa_e=1$) cases are compared. 
    Units are $a^2\omega_p$.} 
  \label{tab:hd_kappa}
\end{table}

\subsection{High-$Z$ Impurity} 
\label{sec:hiz}

%
\begin{figure}
  \includegraphics[width=3in]{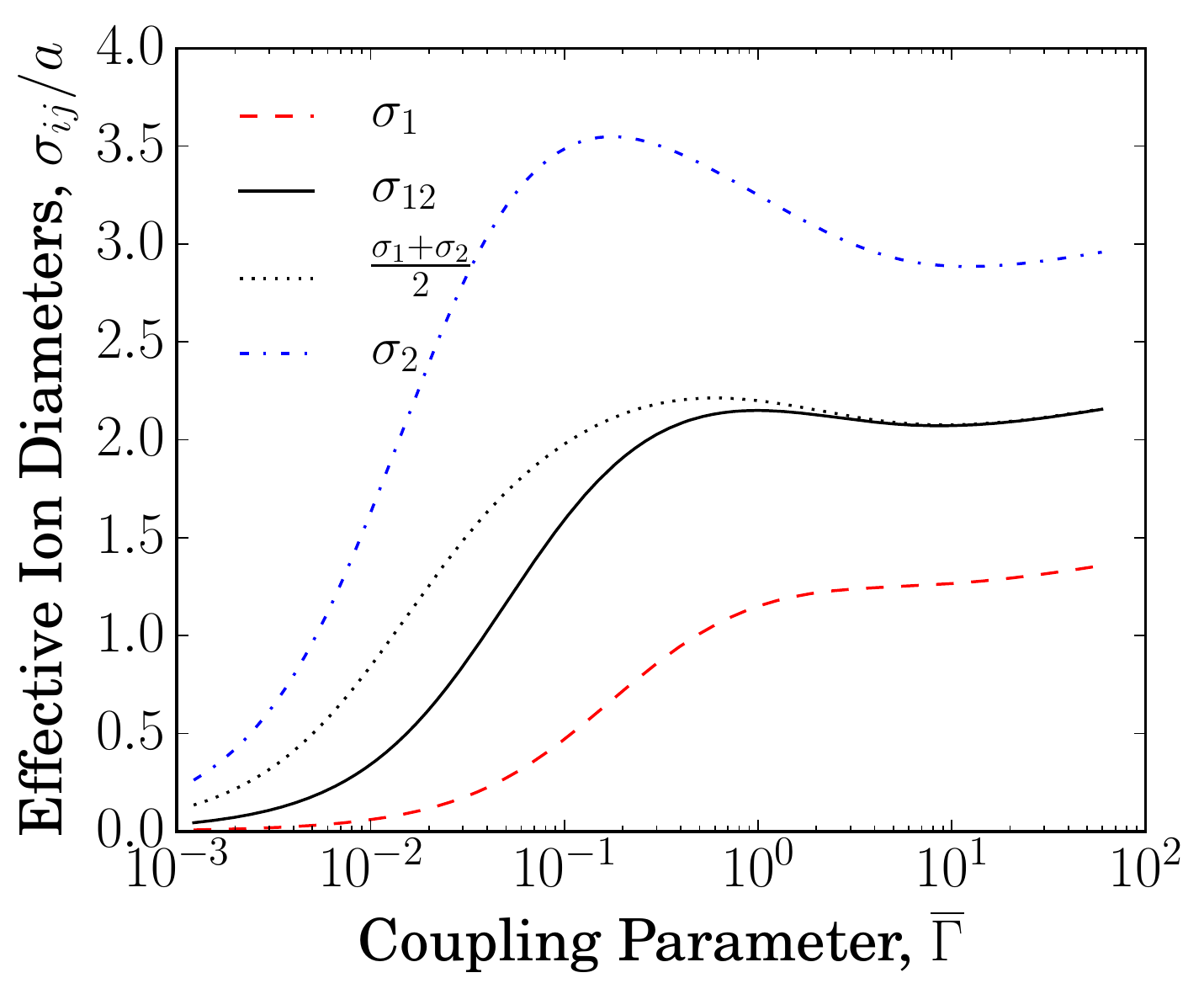}
  \includegraphics[width=3in]{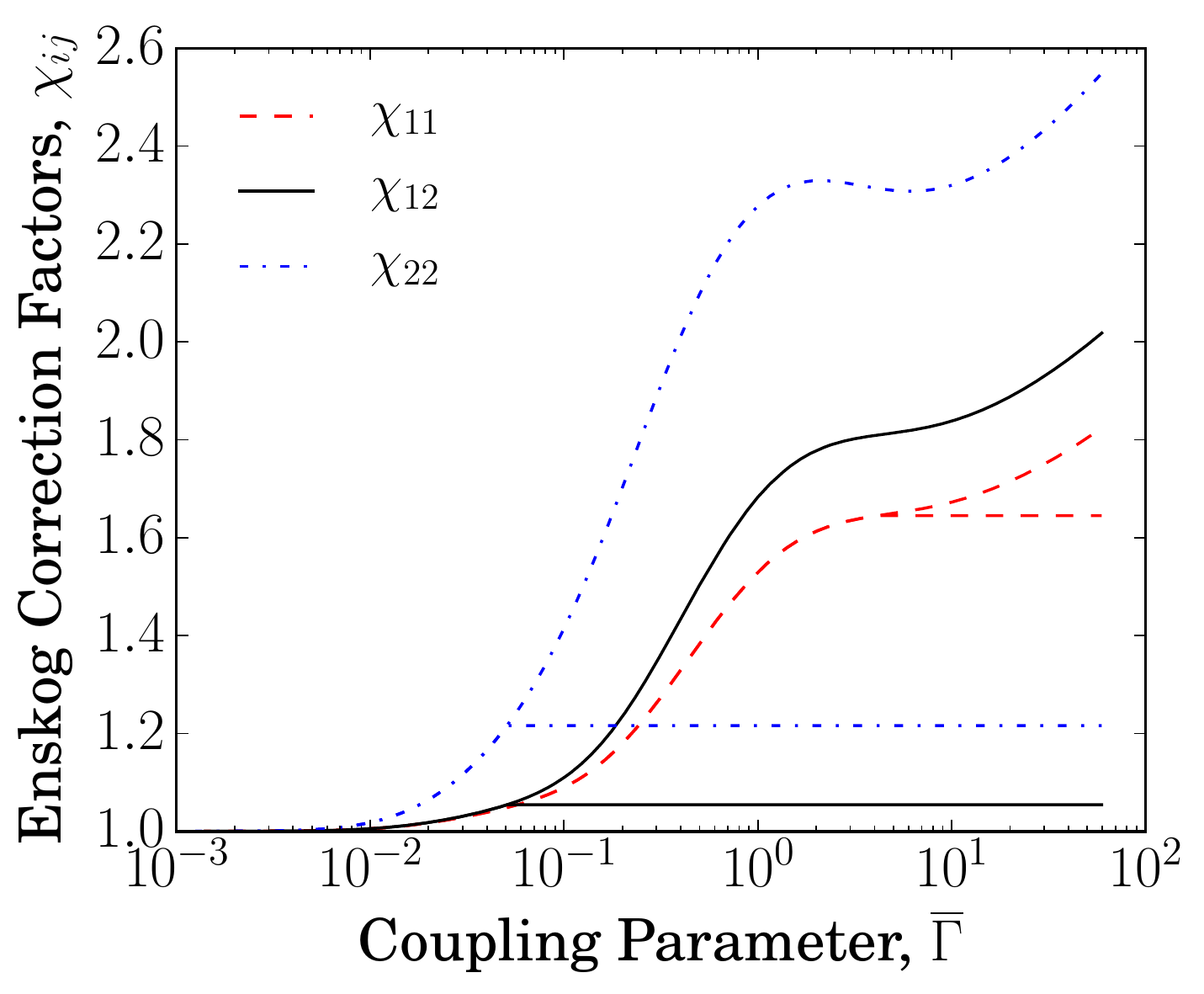}
  \caption{(a) Effective ion diameters and (b) modified Enskog correction factors (both with and without virial cutoff) in a 0.9\% carbon (by mole) \HC BIM\@.}
  \label{fig:HC_y2_100_enskog}
\end{figure}
\begin{figure}
  \centering
  \includegraphics[width=3in]{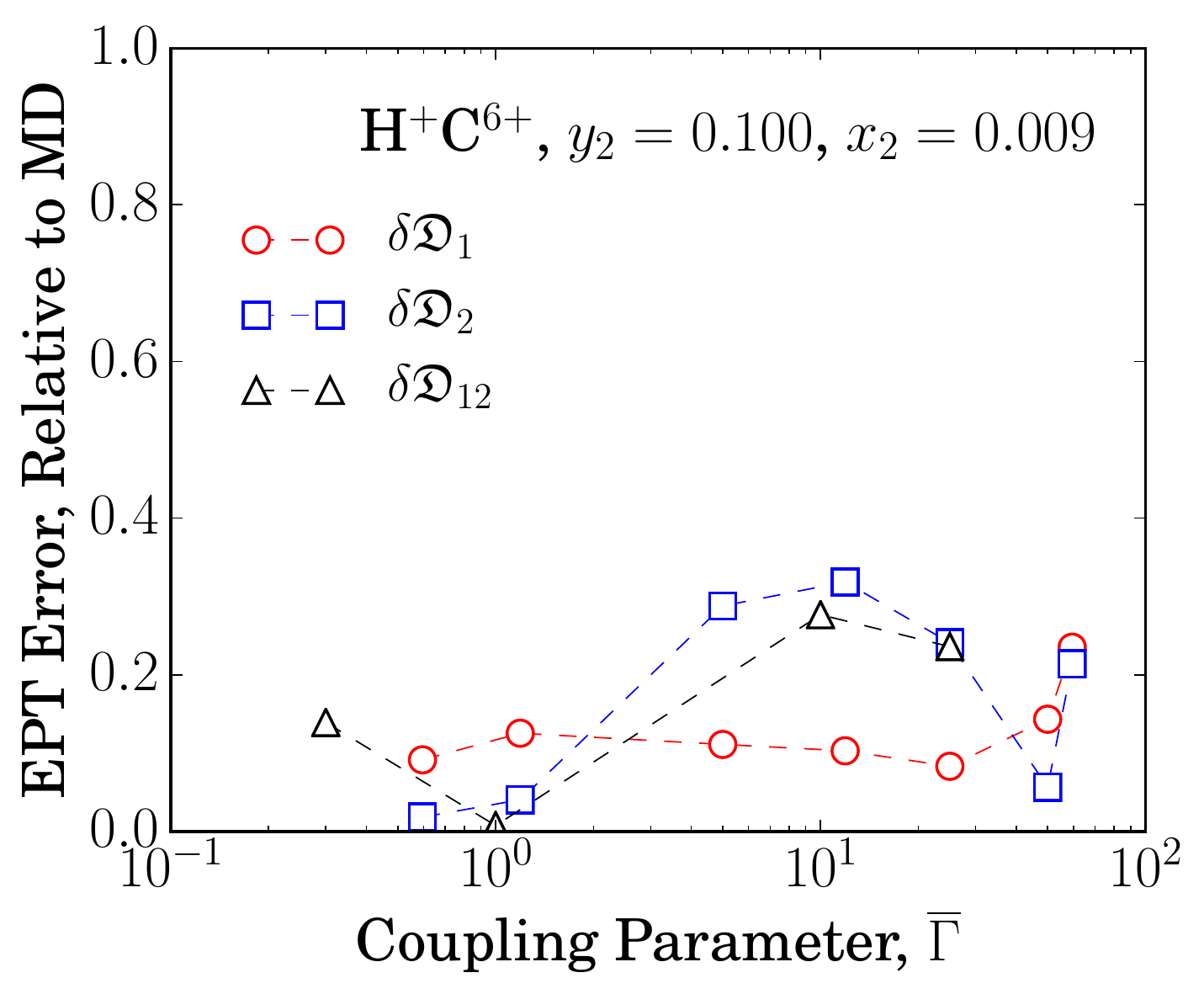}
  \caption{Errors in EPT diffusion coefficients, relative to MD values for a 10\% carbon (by mass) \HC BIM\@. Red circles: hydrogen self-diffusion coefficient. Blue squares: carbon self-diffusion coefficient. Black triangles: interdiffusion coefficient. Dashed lines are included to guide the eye.}
  \label{fig:HC_y2_100_err}
\end{figure}

We classify a BIM as having a high-$Z$ impurity if the more highly charged species is much less abundant than the lesser-charged species.
Diffusion coefficients for an example \HC BIM are plotted in Figures~\ref{fig:bigfig}c~and~f, and effective diameters and modified Enskog corrections are plotted in Figures~\ref{fig:HC_y2_100_enskog}a~and~b. 

In Figure~\ref{fig:HC_y2_100_err}, we plot the relative differences between EPT and MD diffusion coefficients for a \HC with trace abundance of carbon.
EPT predicts the MD interdiffusion coefficient and impurity self-diffusion coefficient within $40\%$ across all coupling strengths studied.
In addition, EPT predicts the majority species self-diffusion coefficient within $20\%$ up to $\gmbar=25$, in line with the range of accuracy observed in the other types of mixtures examined in this work.

\begin{figure}[t!]
  \centering
  \includegraphics[width=3in]{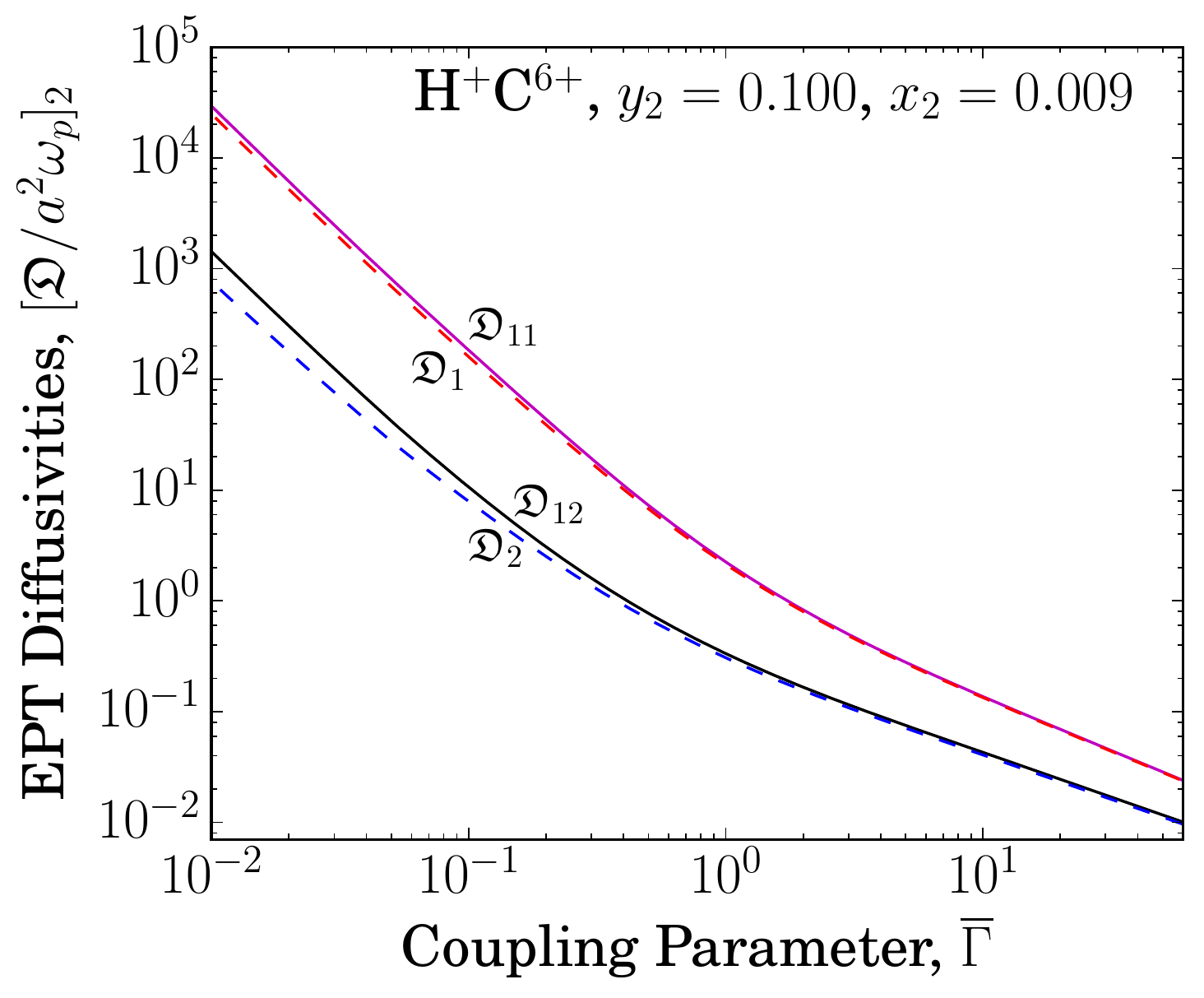}
  \caption{Test of the impurity-limit approximations Eqs.~\eqref{eq:mid12d2}-\eqref{eq:mid11d1} in a 0.9\% (by mole) \HC BIM\@. All diffusion coefficients plotted are second-order EPT results.}
  \label{fig:HC-limit}
\end{figure}

In the impurity limit, EPT predicts 
\begin{eqnarray}
  & \mathfrak{D}_2 \approx \mathfrak{D}_{12}   \label{eq:mid12d2} \\
  & \mathfrak{D}_1 \approx \mathfrak{D}_{11}   \label{eq:mid11d1} \, 
\end{eqnarray}
which follow from Eq.~\eqref{eq:dice} when $x_1 \gg x_2$ and the concentration dependence of $\mathfrak D_{ij}$ is neglected.
These approximations are tested in Figure~\ref{fig:HC-limit}.
At weak coupling, the concentration dependence of $\ordtwo{\mathfrak D_{ij}}$ is non-negligible, and Eqs.~\eqref{eq:mid12d2}-\eqref{eq:mid11d1} do not hold.
However, they become increasingly correct at stronger coupling.
These approximate limits could be useful in molecular dynamics contexts, where self-diffusion coefficients are much less onerous to compute than interdiffusion coefficients.
They may also be of use in developing specialized approximate models of diffusion for this important class of mixtures.
%

\section{Approximations to the Interdiffusion Coefficient}
\label{sec:approx}

In this section, we briefly consider approximate expressions for the interdiffusion coefficient of a binary mixture in terms of the two self-diffusion coefficients.
This is of special practical utility for molecular dynamics studies of diffusion, where the self-diffusion coefficients can be calculated much more expediently than the interdiffusion coefficient, which incurs large computational cost to achieve good statistics.

We focus our attention on the Darken approximation,
\begin{equation}
  \label{eq:darken}
  \mathfrak{D}^{\mathrm{D}}_{12} = x_2 \mathfrak{D}_1 + x_1\mathfrak{D}_2~,
\end{equation}
which is based on a diffusion model where the flux of each species is governed by its self-diffusion coefficient\cite{DarkenAIME1948}.
At the microscopic level, it can be obtained from Eq.~\eqref{eq:gkinter} by neglecting all velocity cross-correlations between different particles.
Another mixing model, the Common Force Model, has been the subject of recent study in plasma physics\cite{HaxhimaliPRE2014}, but for the mixtures shown in Figure~\ref{fig:bigfig}, it did not differ dramatically from the Darken approximation.

We evaluate the interdiffusion coefficient from Eq.~\eqref{eq:darken} using MD self-diffusion coefficients as input.
Comparisons with EPT and MD results for $\mathfrak D_{12}$ are plotted in Figures~\ref{fig:bigfig}d-f.
Relative errors are tabulated in Table~\ref{tab:darken_cfm_err}.
In the case of an ordinary \HHe BIM, we find that the Darken approximation underestimates the interdiffusion coefficient by no worse than $20\%$ up to $\gmbar=25$, consistent with the findings of Hansen et al\cite{HansenPA1985}.
In the case of a high-$Z$ impurity, it is especially successful, agreeing with the Eq.~\eqref{eq:gkinter} within $5\%$ over the entire range of $\gmbar$ examined.
This can be explained by the fact that Eq.~\ref{eq:gkinter} analytically reduces to $\mathfrak D_{12} = \mathfrak D_1$ in the $x_2\to 0$ limit.

While the Darken model is fairly accurate for all cases considered here, it's greatest practical utility seems to be for strongly coupled mixtures containing an impurity.
MD calculations using the Darken model may be used to obtain accurate interdiffusion coefficients for this regime at reduced computational cost.
\begin{table}[h]
  \centering
  \begin{tabular}{c|ccc|ccc|ccc}
    \hline\hline
    \multicolumn{1}{c}{BIM} & \multicolumn{3}{c}{{\HHe}} &  \multicolumn{3}{c}{{\HD}} & \multicolumn{3}{c}{{\HC}} \\ \hline
    $\gmbar$                             & 1     & 10   & 25   & 1     & 10    & 25     & 1    & 10   & 25  \\ \hline
    $\delta\mathfrak D_{12}^{\mathrm{D}}$   & 13.5 & 19.1 & 15.7 &  8.75 &  9.64 &  0.789 & 0.55 & 1.27 & 2.88      \\ 
  \end{tabular}
  \caption{Percent errors in the Darken approximations to $\mathfrak D_{12}$ computed using self-diffusion coefficients from MD simulations. Errors are relative to the value of $\mathfrak D_{12}$ computed directly from MD via Eq.~\eqref{eq:gkinter}. BIM compositions are the same as those in Figure~\ref{fig:bigfig}.}
  \label{tab:darken_cfm_err}
\end{table}

\section{Conclusions}
\label{sec:conc}

Through comparison with molecular dynamics simulations, we have presented evidence that Effective Potential Theory extends weakly coupled plasma theory into the strong coupling regime for the self-diffusion and interdiffusion coefficients of binary ionic mixtures with an accuracy that is sufficient for many current applications.
It improves substantially upon traditional plasma theory, while also being much more practical than molecular dynamics for canvassing the wide BIM parameter space.
Since it is efficient to evaluate, EPT offers a way to build look-up tables for use in fluid simulations of strongly coupled plasmas.
Furthermore, since it requires only the radial distribution functions as input, EPT can be coupled to models for dense plasmas that are more realistic for particular applications than the classical BIM, as was shown in ref.~\cite{DaligaultPRL2016}.
In this way, it is suitable for application to present-day diffusion challenges such as those arising in inertial confinement fusion and stellar astrophysics.

\begin{acknowledgments}
  This material is based upon work supported by the National Science Foundation under Grant No.~PHY-1453736. 
  The work of J.~D.~was supported by Los Alamos National Laboratory LDRD Grant No.~20150520ER. 
\end{acknowledgments}

\appendix
 
\section{Second-Order Corrections to $\mathfrak{D}_{ij}$}
\label{sec:deltaij}

The second-order correction to $\mathfrak{D}_{12}$ is typically given in standard texts in terms of dimensional $\Omega$-integrals\cite{ChapmanCowling,FerzigerKaper}. 
Here we expose the mass and charge dependence by writing it in terms of the dimensionless $\Xi_{ij}^\lk$ defined in Eq.~\eqref{eq:xilk}.
The correction $\Delta_{12}$ is given by
\begin{align}
  \Delta_{12} = 5(\mathrm{C} - 1)^2 \frac{x_1^2 \mathrm P_1 + x_1x_2 \mathrm P_{12} + x_2^2 \mathrm P_2}{x_1^2 \mathrm Q_1 + x_1x_2 \mathrm Q_{12} + x_2^2 \mathrm Q_2} \, ,
\end{align}
where 
\begin{align}
  &\mathrm P_1 = \left(\frac{m_1}{m_1+m_2}\right)^3 \mathrm{E}_1 \\
  &\mathrm P_{12} = 3 \left(\frac{m_1-m_2}{m_1+m_2}\right)^2 + 4\frac{m_{12}}{m_1+m_2} \mathrm A\\
  &\mathrm Q_1 = \mathrm{E}_1 \frac{6 m_1 m_2^2 + 8 m_1^2m_2 \mathrm{A} + m_1^3 (5-4\mathrm {B})}{(m_1+m_2)^3}\\
  &\mathrm Q_{12} = 2\frac{m_{12}}{m_1+m_2} \mathrm E_1 \mathrm E_2 + 3 \left(\frac{m_1-m_2}{m_1+m_2}\right)^2 (5-4\mathrm B)\nonumber \\ 
              &\qquad + \frac{4m_{12}}{m_1+m_2} \mathrm A (11-4\mathrm B)~,
\end{align}
and
\begin{align}
  &\mathrm{A} = \frac{\Xi_{12}^{(2,2)}}{5 \Xi_{12}^{(1,1)}}\\
  &\mathrm{B} = \frac{5\Xi_{12}^{(1,2)} - \Xi_{12}^{(1,3)}}{5 \Xi_{12}^{(1,1)}}\\ &\mathrm{C} = \frac{2 \Xi_{12}^{(1,2)}}{5 \Xi_{12}^{(1,1)}} \\
  &\mathrm{E}_1 = 2 \frac{Z_1^2 m_{2}}{Z_2^2 m_{12}} \frac{\Xi_{11}^{(2,2)}}{\Xi_{12}^{(1,1)}}~,
\end{align}
and $\mathrm{P}_2$, $\mathrm{Q}_2$, and $\mathrm{E}_2$ are obtained by switching the species labels in $\mathrm{P}_1$, $\mathrm{Q}_1$, and $\mathrm{E}_1$ respectively. 
Corrections for $\mathfrak{D}_{ii}$ can be obtained by setting the species labels to be the same, resulting in
\begin{equation}
  \Delta_{ii} =  \frac{(2 \Xi_{ii}^{(1,2)} - 5 \Xi_{ii}^{(1,1)})^2 / \Xi_{ii}^{(1,1)}}{55 \Xi_{ii}^{(1,1)} - 20 \Xi_{ii}^{(1,2)} + 4 \Xi_{ii}^{(1,3)} + 8 \Xi_{ii}^{(2,2)}}~,
\end{equation}
which is of the same form as the OCP case\cite{BaalrudPOP2014}.

\bibliography{refs}

\begin{thebibliography}{34}%
\makeatletter
\providecommand \@ifxundefined [1]{%
 \@ifx{#1\undefined}
}%
\providecommand \@ifnum [1]{%
 \ifnum #1\expandafter \@firstoftwo
 \else \expandafter \@secondoftwo
 \fi
}%
\providecommand \@ifx [1]{%
 \ifx #1\expandafter \@firstoftwo
 \else \expandafter \@secondoftwo
 \fi
}%
\providecommand \natexlab [1]{#1}%
\providecommand \enquote  [1]{``#1''}%
\providecommand \bibnamefont  [1]{#1}%
\providecommand \bibfnamefont [1]{#1}%
\providecommand \citenamefont [1]{#1}%
\providecommand \href@noop [0]{\@secondoftwo}%
\providecommand \href [0]{\begingroup \@sanitize@url \@href}%
\providecommand \@href[1]{\@@startlink{#1}\@@href}%
\providecommand \@@href[1]{\endgroup#1\@@endlink}%
\providecommand \@sanitize@url [0]{\catcode `\\12\catcode `\$12\catcode
  `\&12\catcode `\#12\catcode `\^12\catcode `\_12\catcode `\%12\relax}%
\providecommand \@@startlink[1]{}%
\providecommand \@@endlink[0]{}%
\providecommand \url  [0]{\begingroup\@sanitize@url \@url }%
\providecommand \@url [1]{\endgroup\@href {#1}{\urlprefix }}%
\providecommand \urlprefix  [0]{URL }%
\providecommand \Eprint [0]{\href }%
\providecommand \doibase [0]{http://dx.doi.org/}%
\providecommand \selectlanguage [0]{\@gobble}%
\providecommand \bibinfo  [0]{\@secondoftwo}%
\providecommand \bibfield  [0]{\@secondoftwo}%
\providecommand \translation [1]{[#1]}%
\providecommand \BibitemOpen [0]{}%
\providecommand \bibitemStop [0]{}%
\providecommand \bibitemNoStop [0]{.\EOS\space}%
\providecommand \EOS [0]{\spacefactor3000\relax}%
\providecommand \BibitemShut  [1]{\csname bibitem#1\endcsname}%
\let\auto@bib@innerbib\@empty
\bibitem [{\citenamefont {Hu}\ \emph {et~al.}(2010)\citenamefont {Hu},
  \citenamefont {Militzer}, \citenamefont {Goncharov},\ and\ \citenamefont
  {Skupsky}}]{HuPRL2010}%
  \BibitemOpen
  \bibfield  {author} {\bibinfo {author} {\bibfnamefont {S.~X.}\ \bibnamefont
  {Hu}}, \bibinfo {author} {\bibfnamefont {B.}~\bibnamefont {Militzer}},
  \bibinfo {author} {\bibfnamefont {V.~N.}\ \bibnamefont {Goncharov}}, \ and\
  \bibinfo {author} {\bibfnamefont {S.}~\bibnamefont {Skupsky}},\ }\href
  {\doibase 10.1103/PhysRevLett.104.235003} {\bibfield  {journal} {\bibinfo
  {journal} {\prl}\ }\textbf {\bibinfo {volume} {104}},\ \bibinfo {pages}
  {235003} (\bibinfo {year} {2010})}\BibitemShut {NoStop}%
\bibitem [{\citenamefont {Ikezi}(1986)}]{IkeziPF1986}%
  \BibitemOpen
  \bibfield  {author} {\bibinfo {author} {\bibfnamefont {H.}~\bibnamefont
  {Ikezi}},\ }\href {\doibase http://dx.doi.org/10.1063/1.865653} {\bibfield
  {journal} {\bibinfo  {journal} {Physics of Fluids}\ }\textbf {\bibinfo
  {volume} {29}},\ \bibinfo {pages} {1764} (\bibinfo {year}
  {1986})}\BibitemShut {NoStop}%
\bibitem [{\citenamefont {{Killian}}\ \emph {et~al.}(1999)\citenamefont
  {{Killian}}, \citenamefont {{Kulin}}, \citenamefont {{Bergeson}},
  \citenamefont {{Orozco}}, \citenamefont {{Orzel}},\ and\ \citenamefont
  {{Rolston}}}]{KillianPRL1999}%
  \BibitemOpen
  \bibfield  {author} {\bibinfo {author} {\bibfnamefont {T.~C.}\ \bibnamefont
  {{Killian}}}, \bibinfo {author} {\bibfnamefont {S.}~\bibnamefont {{Kulin}}},
  \bibinfo {author} {\bibfnamefont {S.~D.}\ \bibnamefont {{Bergeson}}},
  \bibinfo {author} {\bibfnamefont {L.~A.}\ \bibnamefont {{Orozco}}}, \bibinfo
  {author} {\bibfnamefont {C.}~\bibnamefont {{Orzel}}}, \ and\ \bibinfo
  {author} {\bibfnamefont {S.~L.}\ \bibnamefont {{Rolston}}},\ }\href {\doibase
  10.1103/PhysRevLett.83.4776} {\bibfield  {journal} {\bibinfo  {journal}
  {\prl}\ }\textbf {\bibinfo {volume} {83}},\ \bibinfo {pages} {4776} (\bibinfo
  {year} {1999})},\ \Eprint {http://arxiv.org/abs/physics/9908051}
  {physics/9908051} \BibitemShut {NoStop}%
\bibitem [{\citenamefont {Strickler}\ \emph {et~al.}(2016)\citenamefont
  {Strickler}, \citenamefont {Langin}, \citenamefont {McQuillen}, \citenamefont
  {Daligault},\ and\ \citenamefont {Killian}}]{StricklerPRX2016}%
  \BibitemOpen
  \bibfield  {author} {\bibinfo {author} {\bibfnamefont {T.~S.}\ \bibnamefont
  {Strickler}}, \bibinfo {author} {\bibfnamefont {T.~K.}\ \bibnamefont
  {Langin}}, \bibinfo {author} {\bibfnamefont {P.}~\bibnamefont {McQuillen}},
  \bibinfo {author} {\bibfnamefont {J.}~\bibnamefont {Daligault}}, \ and\
  \bibinfo {author} {\bibfnamefont {T.~C.}\ \bibnamefont {Killian}},\ }\href
  {\doibase 10.1103/PhysRevX.6.021021} {\bibfield  {journal} {\bibinfo
  {journal} {Phys. Rev. X}\ }\textbf {\bibinfo {volume} {6}},\ \bibinfo {pages}
  {021021} (\bibinfo {year} {2016})}\BibitemShut {NoStop}%
\bibitem [{\citenamefont {{Bollinger}}\ and\ \citenamefont
  {{Wineland}}(1984)}]{BollingerPRL1984}%
  \BibitemOpen
  \bibfield  {author} {\bibinfo {author} {\bibfnamefont {J.~J.}\ \bibnamefont
  {{Bollinger}}}\ and\ \bibinfo {author} {\bibfnamefont {D.~J.}\ \bibnamefont
  {{Wineland}}},\ }\href {\doibase 10.1103/PhysRevLett.53.348} {\bibfield
  {journal} {\bibinfo  {journal} {\prl}\ }\textbf {\bibinfo {volume} {53}},\
  \bibinfo {pages} {348} (\bibinfo {year} {1984})}\BibitemShut {NoStop}%
\bibitem [{\citenamefont {{Paquette}}\ \emph
  {et~al.}(1986{\natexlab{a}})\citenamefont {{Paquette}}, \citenamefont
  {{Pelletier}}, \citenamefont {{Fontaine}},\ and\ \citenamefont
  {{Michaud}}}]{PaquetteAPJS1986b}%
  \BibitemOpen
  \bibfield  {author} {\bibinfo {author} {\bibfnamefont {C.}~\bibnamefont
  {{Paquette}}}, \bibinfo {author} {\bibfnamefont {C.}~\bibnamefont
  {{Pelletier}}}, \bibinfo {author} {\bibfnamefont {G.}~\bibnamefont
  {{Fontaine}}}, \ and\ \bibinfo {author} {\bibfnamefont {G.}~\bibnamefont
  {{Michaud}}},\ }\href {\doibase 10.1086/191112} {\bibfield  {journal}
  {\bibinfo  {journal} {Astrophy. J. Suppl.}\ }\textbf {\bibinfo {volume}
  {61}},\ \bibinfo {pages} {197} (\bibinfo {year}
  {1986}{\natexlab{a}})}\BibitemShut {NoStop}%
\bibitem [{\citenamefont {{Beznogov}}\ and\ \citenamefont
  {{Yakovlev}}(2013)}]{BeznogovPRL2013}%
  \BibitemOpen
  \bibfield  {author} {\bibinfo {author} {\bibfnamefont {M.~V.}\ \bibnamefont
  {{Beznogov}}}\ and\ \bibinfo {author} {\bibfnamefont {D.~G.}\ \bibnamefont
  {{Yakovlev}}},\ }\href {\doibase 10.1103/PhysRevLett.111.161101} {\bibfield
  {journal} {\bibinfo  {journal} {\prl}\ }\textbf {\bibinfo {volume} {111}},\
  \bibinfo {eid} {161101} (\bibinfo {year} {2013})},\ \Eprint
  {http://arxiv.org/abs/1307.6060} {arXiv:1307.6060 [astro-ph.SR]} \BibitemShut
  {NoStop}%
\bibitem [{\citenamefont {{Hansen}}\ \emph {et~al.}(1985)\citenamefont
  {{Hansen}}, \citenamefont {{Joly}},\ and\ \citenamefont
  {{McDonald}}}]{HansenPA1985}%
  \BibitemOpen
  \bibfield  {author} {\bibinfo {author} {\bibfnamefont {J.~P.}\ \bibnamefont
  {{Hansen}}}, \bibinfo {author} {\bibfnamefont {F.}~\bibnamefont {{Joly}}}, \
  and\ \bibinfo {author} {\bibfnamefont {I.~R.}\ \bibnamefont {{McDonald}}},\
  }\href {\doibase 10.1016/0378-4371(85)90022-6} {\bibfield  {journal}
  {\bibinfo  {journal} {Physica A}\ }\textbf {\bibinfo {volume} {132}},\
  \bibinfo {pages} {472} (\bibinfo {year} {1985})}\BibitemShut {NoStop}%
\bibitem [{\citenamefont {{Daligault}}(2012)}]{DaligaultPRL2012}%
  \BibitemOpen
  \bibfield  {author} {\bibinfo {author} {\bibfnamefont {J.}~\bibnamefont
  {{Daligault}}},\ }\href {\doibase 10.1103/PhysRevLett.108.225004} {\bibfield
  {journal} {\bibinfo  {journal} {\prl}\ }\textbf {\bibinfo {volume} {108}},\
  \bibinfo {eid} {225004} (\bibinfo {year} {2012})}\BibitemShut {NoStop}%
\bibitem [{\citenamefont {{Haxhimali}}\ \emph {et~al.}(2014)\citenamefont
  {{Haxhimali}}, \citenamefont {{Rudd}}, \citenamefont {{Cabot}},\ and\
  \citenamefont {{Graziani}}}]{HaxhimaliPRE2014}%
  \BibitemOpen
  \bibfield  {author} {\bibinfo {author} {\bibfnamefont {T.}~\bibnamefont
  {{Haxhimali}}}, \bibinfo {author} {\bibfnamefont {R.~E.}\ \bibnamefont
  {{Rudd}}}, \bibinfo {author} {\bibfnamefont {W.~H.}\ \bibnamefont {{Cabot}}},
  \ and\ \bibinfo {author} {\bibfnamefont {F.~R.}\ \bibnamefont {{Graziani}}},\
  }\href {\doibase 10.1103/PhysRevE.90.023104} {\bibfield  {journal} {\bibinfo
  {journal} {\pre}\ }\textbf {\bibinfo {volume} {90}},\ \bibinfo {eid} {023104}
  (\bibinfo {year} {2014})}\BibitemShut {NoStop}%
\bibitem [{\citenamefont {{Liboff}}(1959)}]{LiboffPF1959}%
  \BibitemOpen
  \bibfield  {author} {\bibinfo {author} {\bibfnamefont {R.~L.}\ \bibnamefont
  {{Liboff}}},\ }\href {\doibase 10.1063/1.1724389} {\bibfield  {journal}
  {\bibinfo  {journal} {Phys. Fluids}\ }\textbf {\bibinfo {volume} {2}},\
  \bibinfo {pages} {40} (\bibinfo {year} {1959})}\BibitemShut {NoStop}%
\bibitem [{\citenamefont {{Paquette}}\ \emph
  {et~al.}(1986{\natexlab{b}})\citenamefont {{Paquette}}, \citenamefont
  {{Pelletier}}, \citenamefont {{Fontaine}},\ and\ \citenamefont
  {{Michaud}}}]{PaquetteAPJS1986a}%
  \BibitemOpen
  \bibfield  {author} {\bibinfo {author} {\bibfnamefont {C.}~\bibnamefont
  {{Paquette}}}, \bibinfo {author} {\bibfnamefont {C.}~\bibnamefont
  {{Pelletier}}}, \bibinfo {author} {\bibfnamefont {G.}~\bibnamefont
  {{Fontaine}}}, \ and\ \bibinfo {author} {\bibfnamefont {G.}~\bibnamefont
  {{Michaud}}},\ }\href {\doibase 10.1086/191111} {\bibfield  {journal}
  {\bibinfo  {journal} {Astrophys. J. Suppl.}\ }\textbf {\bibinfo {volume}
  {61}},\ \bibinfo {pages} {177} (\bibinfo {year}
  {1986}{\natexlab{b}})}\BibitemShut {NoStop}%
\bibitem [{\citenamefont {{Baalrud}}\ and\ \citenamefont
  {{Daligault}}(2013)}]{BaalrudPRL2013}%
  \BibitemOpen
  \bibfield  {author} {\bibinfo {author} {\bibfnamefont {S.~D.}\ \bibnamefont
  {{Baalrud}}}\ and\ \bibinfo {author} {\bibfnamefont {J.}~\bibnamefont
  {{Daligault}}},\ }\href {\doibase 10.1103/PhysRevLett.110.235001} {\bibfield
  {journal} {\bibinfo  {journal} {\prl}\ }\textbf {\bibinfo {volume} {110}},\
  \bibinfo {eid} {235001} (\bibinfo {year} {2013})},\ \Eprint
  {http://arxiv.org/abs/1303.3202} {arXiv:1303.3202 [physics.plasm-ph]}
  \BibitemShut {NoStop}%
\bibitem [{\citenamefont {{Daligault}}\ \emph {et~al.}(2016)\citenamefont
  {{Daligault}}, \citenamefont {{Baalrud}}, \citenamefont {{Starrett}},
  \citenamefont {{Saumon}},\ and\ \citenamefont
  {{Sjostrom}}}]{DaligaultPRL2016}%
  \BibitemOpen
  \bibfield  {author} {\bibinfo {author} {\bibfnamefont {J.}~\bibnamefont
  {{Daligault}}}, \bibinfo {author} {\bibfnamefont {S.~D.}\ \bibnamefont
  {{Baalrud}}}, \bibinfo {author} {\bibfnamefont {C.~E.}\ \bibnamefont
  {{Starrett}}}, \bibinfo {author} {\bibfnamefont {D.}~\bibnamefont
  {{Saumon}}}, \ and\ \bibinfo {author} {\bibfnamefont {T.}~\bibnamefont
  {{Sjostrom}}},\ }\href {\doibase 10.1103/PhysRevLett.116.075002} {\bibfield
  {journal} {\bibinfo  {journal} {\prl}\ }\textbf {\bibinfo {volume} {116}},\
  \bibinfo {eid} {075002} (\bibinfo {year} {2016})}\BibitemShut {NoStop}%
\bibitem [{\citenamefont {{Baalrud}}\ and\ \citenamefont
  {{Daligault}}(2015)}]{BaalrudPRE2015}%
  \BibitemOpen
  \bibfield  {author} {\bibinfo {author} {\bibfnamefont {S.~D.}\ \bibnamefont
  {{Baalrud}}}\ and\ \bibinfo {author} {\bibfnamefont {J.}~\bibnamefont
  {{Daligault}}},\ }\href {\doibase 10.1103/PhysRevE.91.063107} {\bibfield
  {journal} {\bibinfo  {journal} {\pre}\ }\textbf {\bibinfo {volume} {91}},\
  \bibinfo {eid} {063107} (\bibinfo {year} {2015})},\ \Eprint
  {http://arxiv.org/abs/1506.03112} {arXiv:1506.03112 [physics.plasm-ph]}
  \BibitemShut {NoStop}%
\bibitem [{\citenamefont {{Donk{\'o}}}\ \emph {et~al.}(2002)\citenamefont
  {{Donk{\'o}}}, \citenamefont {{Kalman}},\ and\ \citenamefont
  {{Golden}}}]{DonkoPRL2002}%
  \BibitemOpen
  \bibfield  {author} {\bibinfo {author} {\bibfnamefont {Z.}~\bibnamefont
  {{Donk{\'o}}}}, \bibinfo {author} {\bibfnamefont {G.~J.}\ \bibnamefont
  {{Kalman}}}, \ and\ \bibinfo {author} {\bibfnamefont {K.~I.}\ \bibnamefont
  {{Golden}}},\ }\href {\doibase 10.1103/PhysRevLett.88.225001} {\bibfield
  {journal} {\bibinfo  {journal} {\prl}\ }\textbf {\bibinfo {volume} {88}},\
  \bibinfo {eid} {225001} (\bibinfo {year} {2002})}\BibitemShut {NoStop}%
\bibitem [{\citenamefont {{Beznogov}}\ and\ \citenamefont
  {{Yakovlev}}(2014)}]{BeznogovPRE2014}%
  \BibitemOpen
  \bibfield  {author} {\bibinfo {author} {\bibfnamefont {M.~V.}\ \bibnamefont
  {{Beznogov}}}\ and\ \bibinfo {author} {\bibfnamefont {D.~G.}\ \bibnamefont
  {{Yakovlev}}},\ }\href {\doibase 10.1103/PhysRevE.90.033102} {\bibfield
  {journal} {\bibinfo  {journal} {\pre}\ }\textbf {\bibinfo {volume} {90}},\
  \bibinfo {eid} {033102} (\bibinfo {year} {2014})},\ \Eprint
  {http://arxiv.org/abs/1409.1407} {arXiv:1409.1407 [astro-ph.SR]} \BibitemShut
  {NoStop}%
\bibitem [{\citenamefont {{Hansen}}\ and\ \citenamefont
  {{Vieillefosse}}(1976)}]{HansenPRL1976}%
  \BibitemOpen
  \bibfield  {author} {\bibinfo {author} {\bibfnamefont {J.-P.}\ \bibnamefont
  {{Hansen}}}\ and\ \bibinfo {author} {\bibfnamefont {P.}~\bibnamefont
  {{Vieillefosse}}},\ }\href {\doibase 10.1103/PhysRevLett.37.391} {\bibfield
  {journal} {\bibinfo  {journal} {\prl}\ }\textbf {\bibinfo {volume} {37}},\
  \bibinfo {pages} {391} (\bibinfo {year} {1976})}\BibitemShut {NoStop}%
\bibitem [{\citenamefont {{Baalrud}}\ and\ \citenamefont
  {{Daligault}}(2014)}]{BaalrudPOP2014}%
  \BibitemOpen
  \bibfield  {author} {\bibinfo {author} {\bibfnamefont {S.~D.}\ \bibnamefont
  {{Baalrud}}}\ and\ \bibinfo {author} {\bibfnamefont {J.}~\bibnamefont
  {{Daligault}}},\ }\href {\doibase 10.1063/1.4875282} {\bibfield  {journal}
  {\bibinfo  {journal} {Phys. Plasmas}\ }\textbf {\bibinfo {volume} {21}},\
  \bibinfo {eid} {055707} (\bibinfo {year} {2014})},\ \Eprint
  {http://arxiv.org/abs/1403.1882} {arXiv:1403.1882 [physics.plasm-ph]}
  \BibitemShut {NoStop}%
\bibitem [{\citenamefont {{Chapman}}\ and\ \citenamefont
  {{Cowling}}(1970)}]{ChapmanCowling}%
  \BibitemOpen
  \bibfield  {author} {\bibinfo {author} {\bibfnamefont {S.}~\bibnamefont
  {{Chapman}}}\ and\ \bibinfo {author} {\bibfnamefont {T.~G.}\ \bibnamefont
  {{Cowling}}},\ }\href@noop {} {\emph {\bibinfo {title} {{The Mathematical
  Theory of Non-Uniform Gases}}}},\ \bibinfo {edition} {3rd}\ ed.\ (\bibinfo
  {publisher} {Cambridge University Press},\ \bibinfo {year}
  {1970})\BibitemShut {NoStop}%
\bibitem [{\citenamefont {{Hansen}}\ and\ \citenamefont
  {{MacDonald}}(1976)}]{HansenMacDonald}%
  \BibitemOpen
  \bibfield  {author} {\bibinfo {author} {\bibfnamefont {J.~P.}\ \bibnamefont
  {{Hansen}}}\ and\ \bibinfo {author} {\bibfnamefont {I.~R.}\ \bibnamefont
  {{MacDonald}}},\ }\href@noop {} {\emph {\bibinfo {title} {{Theory of Simple
  Liquids}}}},\ \bibinfo {edition} {1st}\ ed.\ (\bibinfo  {publisher} {Academic
  Press},\ \bibinfo {year} {1976})\BibitemShut {NoStop}%
\bibitem [{\citenamefont {{Ferziger}}\ and\ \citenamefont
  {{Kaper}}(1972)}]{FerzigerKaper}%
  \BibitemOpen
  \bibfield  {author} {\bibinfo {author} {\bibfnamefont {J.~H.}\ \bibnamefont
  {{Ferziger}}}\ and\ \bibinfo {author} {\bibfnamefont {H.~G.}\ \bibnamefont
  {{Kaper}}},\ }\href@noop {} {\emph {\bibinfo {title} {Mathematical Theory of
  Transport Processes in Gases}}}\ (\bibinfo  {publisher} {North-Holland Pub.
  Co Amsterdam},\ \bibinfo {year} {1972})\BibitemShut {NoStop}%
\bibitem [{\citenamefont {{Enskog}}(1922)}]{EnskogKSVAH1922}%
  \BibitemOpen
  \bibfield  {author} {\bibinfo {author} {\bibfnamefont {D.}~\bibnamefont
  {{Enskog}}},\ }\href@noop {} {\bibfield  {journal} {\bibinfo  {journal} {K.
  Sven. Vet.-Ak. Handl .}\ }\textbf {\bibinfo {volume} {63}} (\bibinfo {year}
  {1922})}\BibitemShut {NoStop}%
\bibitem [{\citenamefont {{Pi{\~n}a}}(1974)}]{PinaJSP1974}%
  \BibitemOpen
  \bibfield  {author} {\bibinfo {author} {\bibfnamefont {E.}~\bibnamefont
  {{Pi{\~n}a}}},\ }\href {\doibase 10.1007/BF01026734} {\bibfield  {journal}
  {\bibinfo  {journal} {Journal of Statistical Physics}\ }\textbf {\bibinfo
  {volume} {11}},\ \bibinfo {pages} {433} (\bibinfo {year} {1974})}\BibitemShut
  {NoStop}%
\bibitem [{Note1()}]{Note1}%
  \BibitemOpen
  \bibinfo {note} {Section 16.6 of Ref.~\cite {ChapmanCowling}}\BibitemShut
  {NoStop}%
\bibitem [{\citenamefont {{Kincaid}}\ \emph {et~al.}(1983)\citenamefont
  {{Kincaid}}, \citenamefont {{L\'opez de Haro}},\ and\ \citenamefont
  {{Cohen}}}]{KincaidJCP1983}%
  \BibitemOpen
  \bibfield  {author} {\bibinfo {author} {\bibfnamefont {J.~M.}\ \bibnamefont
  {{Kincaid}}}, \bibinfo {author} {\bibfnamefont {M.}~\bibnamefont {{L\'opez de
  Haro}}}, \ and\ \bibinfo {author} {\bibfnamefont {E.~G.~D.}\ \bibnamefont
  {{Cohen}}},\ }\href@noop {} {\bibfield  {journal} {\bibinfo  {journal}
  {\jcp}\ }\textbf {\bibinfo {volume} {79}},\ \bibinfo {pages} {4509} (\bibinfo
  {year} {1983})}\BibitemShut {NoStop}%
\bibitem [{\citenamefont {Krishna}\ and\ \citenamefont {van
  Baten}(2005)}]{KrishnaIECR2005}%
  \BibitemOpen
  \bibfield  {author} {\bibinfo {author} {\bibfnamefont {R.}~\bibnamefont
  {Krishna}}\ and\ \bibinfo {author} {\bibfnamefont {J.~M.}\ \bibnamefont {van
  Baten}},\ }\href@noop {} {\bibfield  {journal} {\bibinfo  {journal} {Ind.
  Eng. Chem. Res.}\ }\textbf {\bibinfo {volume} {44}},\ \bibinfo {pages} {6939}
  (\bibinfo {year} {2005})}\BibitemShut {NoStop}%
\bibitem [{\citenamefont {Reif}(2009)}]{Reif}%
  \BibitemOpen
  \bibfield  {author} {\bibinfo {author} {\bibfnamefont {F.}~\bibnamefont
  {Reif}},\ }\href@noop {} {\emph {\bibinfo {title} {Fundamentals of
  Statistical and Thermal Physics}}}\ (\bibinfo  {publisher} {Waveland Press},\
  \bibinfo {year} {2009})\BibitemShut {NoStop}%
\bibitem [{\citenamefont {{Daligault}}()}]{JeromeDraft}%
  \BibitemOpen
  \bibfield  {author} {\bibinfo {author} {\bibfnamefont {J.}~\bibnamefont
  {{Daligault}}},\ }\href@noop {} {\enquote {\bibinfo {title} {{Molecular
  Dynamics Study of Diffuson in Binary Ionic Mixtures (I):~Self-Diffusion}},}\
  }\bibinfo {note} {(Unpubplished)}\BibitemShut {NoStop}%
\bibitem [{\citenamefont {{de~Groot}}\ and\ \citenamefont
  {{Mazur}}(1962)}]{deGrootMazur}%
  \BibitemOpen
  \bibfield  {author} {\bibinfo {author} {\bibfnamefont {S.~R.}\ \bibnamefont
  {{de~Groot}}}\ and\ \bibinfo {author} {\bibfnamefont {P.}~\bibnamefont
  {{Mazur}}},\ }\href@noop {} {\emph {\bibinfo {title} {{Non-Equilibrium
  Thermodynamics}}}}\ (\bibinfo  {publisher} {{North-Holland Pub. Co
  Amsterdam}},\ \bibinfo {year} {1962})\BibitemShut {NoStop}%
\bibitem [{\citenamefont {{Boercker}}\ and\ \citenamefont
  {{Pollock}}(1987)}]{BoerckerPRA1987}%
  \BibitemOpen
  \bibfield  {author} {\bibinfo {author} {\bibfnamefont {D.~B.}\ \bibnamefont
  {{Boercker}}}\ and\ \bibinfo {author} {\bibfnamefont {E.~L.}\ \bibnamefont
  {{Pollock}}},\ }\href {\doibase 10.1103/PhysRevA.36.1779} {\bibfield
  {journal} {\bibinfo  {journal} {\pra}\ }\textbf {\bibinfo {volume} {36}},\
  \bibinfo {pages} {1779} (\bibinfo {year} {1987})}\BibitemShut {NoStop}%
\bibitem [{\citenamefont {{Kincaid}}(1978)}]{KincaidPLA1978}%
  \BibitemOpen
  \bibfield  {author} {\bibinfo {author} {\bibfnamefont {J.~M.}\ \bibnamefont
  {{Kincaid}}},\ }\href {\doibase 10.1016/0375-9601(78)90673-4} {\bibfield
  {journal} {\bibinfo  {journal} {Physics Letters A}\ }\textbf {\bibinfo
  {volume} {64}},\ \bibinfo {pages} {429} (\bibinfo {year} {1978})}\BibitemShut
  {NoStop}%
\bibitem [{\citenamefont {{L{\'o}pez de Haro}}\ \emph
  {et~al.}(1983)\citenamefont {{L{\'o}pez de Haro}}, \citenamefont {{Cohen}},\
  and\ \citenamefont {{Kincaid}}}]{LopezdeHaroJCP1983}%
  \BibitemOpen
  \bibfield  {author} {\bibinfo {author} {\bibfnamefont {M.}~\bibnamefont
  {{L{\'o}pez de Haro}}}, \bibinfo {author} {\bibfnamefont {E.~G.~D.}\
  \bibnamefont {{Cohen}}}, \ and\ \bibinfo {author} {\bibfnamefont {J.~M.}\
  \bibnamefont {{Kincaid}}},\ }\href {\doibase 10.1063/1.444985} {\bibfield
  {journal} {\bibinfo  {journal} {\jcp}\ }\textbf {\bibinfo {volume} {78}},\
  \bibinfo {pages} {2746} (\bibinfo {year} {1983})}\BibitemShut {NoStop}%
\bibitem [{\citenamefont {{Darken}}(1948)}]{DarkenAIME1948}%
  \BibitemOpen
  \bibfield  {author} {\bibinfo {author} {\bibfnamefont {L.~S.}\ \bibnamefont
  {{Darken}}},\ }\href@noop {} {\bibfield  {journal} {\bibinfo  {journal}
  {Trans.~AIME}\ }\textbf {\bibinfo {volume} {175}},\ \bibinfo {pages} {184}
  (\bibinfo {year} {1948})}\BibitemShut {NoStop}%
\end{thebibliography}%

\end{document}